\documentclass[superscriptaddress,twocolumn,10pt,prx,floatfix,nofootinbib,sectionbib,sort&compress]{revtex4-2}

\input{glyphtounicode}
\usepackage[T1]{fontenc}
\usepackage[utf8]{inputenc}
\usepackage[english]{babel}
\usepackage{amsmath}  
\usepackage{amsfonts} 
\usepackage{amssymb}
\usepackage{graphicx} 
\usepackage[pdftex,bookmarks=true,bookmarksopen,bookmarksnumbered,colorlinks,linkcolor=blue,citecolor=blue,urlcolor=blue]{hyperref}
\graphicspath{ {./} }

\usepackage{tabularx}
\usepackage{array}
\usepackage[table]{xcolor}
\usepackage{xcolor}
\usepackage{braket}
\usepackage{booktabs}
\usepackage{multirow}
\usepackage{caption}
\usepackage{hyperref}
\usepackage{dsfont} 

\usepackage[resetlabels,labeled]{multibib}
\usepackage{comment}
\usepackage{siunitx}
\usepackage[shortlabels]{enumitem}

\newcommand{\aref}[1]{\hyperref[#1]{Appendix~\ref*{#1}}}
\DeclareCaptionLabelSeparator{line}{\hspace{2pt}\bf{|}\hspace{2pt}}

\begin{document}

\captionsetup[figure]{name={\bf{Fig.}},labelsep=line,justification=centerlast,font=small}
\renewcommand{\equationautorefname}{Eq.}
\renewcommand{\figureautorefname}{Fig.}
\renewcommand*{\sectionautorefname}{Sec.}

\title{High-fidelity operation and algorithmic initialisation of spin qubits above one kelvin}


\author{Jonathan Y. Huang}
\email{yue.huang6@unsw.edu.au}
\affiliation{School of Electrical Engineering and Telecommunications, University of New South Wales, Sydney, NSW 2052, Australia}
\author{Rocky Y. Su}
\affiliation{School of Electrical Engineering and Telecommunications, University of New South Wales, Sydney, NSW 2052, Australia}
\author{Wee Han Lim}
\affiliation{School of Electrical Engineering and Telecommunications, University of New South Wales, Sydney, NSW 2052, Australia}
\affiliation{Diraq Pty. Ltd., Sydney, NSW, Australia}
\author{MengKe Feng}
\affiliation{School of Electrical Engineering and Telecommunications, University of New South Wales, Sydney, NSW 2052, Australia}
\author{Barnaby van Straaten}
\affiliation{Department of Engineering Science, University of Oxford, Oxford OX1 3PH, United Kingdom}
\author{Brandon Severin}
\affiliation{School of Electrical Engineering and Telecommunications, University of New South Wales, Sydney, NSW 2052, Australia}
\affiliation{Department of Engineering Science, University of Oxford, Oxford OX1 3PH, United Kingdom}
\author{Will Gilbert}
\affiliation{School of Electrical Engineering and Telecommunications, University of New South Wales, Sydney, NSW 2052, Australia}
\affiliation{Diraq Pty. Ltd., Sydney, NSW, Australia}
\author{Nard Dumoulin Stuyck}
\affiliation{School of Electrical Engineering and Telecommunications, University of New South Wales, Sydney, NSW 2052, Australia}
\affiliation{Diraq Pty. Ltd., Sydney, NSW, Australia}
\author{Tuomo Tanttu}
\affiliation{School of Electrical Engineering and Telecommunications, University of New South Wales, Sydney, NSW 2052, Australia}
\affiliation{Diraq Pty. Ltd., Sydney, NSW, Australia}
\author{Santiago Serrano}
\affiliation{School of Electrical Engineering and Telecommunications, University of New South Wales, Sydney, NSW 2052, Australia}
\author{Jesus D. Cifuentes}
\affiliation{School of Electrical Engineering and Telecommunications, University of New South Wales, Sydney, NSW 2052, Australia}
\author{Ingvild Hansen}
\affiliation{School of Electrical Engineering and Telecommunications, University of New South Wales, Sydney, NSW 2052, Australia}
\author{Amanda E. Seedhouse}
\affiliation{School of Electrical Engineering and Telecommunications, University of New South Wales, Sydney, NSW 2052, Australia}
\author{Ensar Vahapoglu}
\affiliation{School of Electrical Engineering and Telecommunications, University of New South Wales, Sydney, NSW 2052, Australia}
\affiliation{Diraq Pty. Ltd., Sydney, NSW, Australia}
\author{Nikolay V. Abrosimov}
\affiliation{Leibniz-Institut für Kristallzüchtung, 12489 Berlin, Germany}
\author{Hans-Joachim Pohl}
\affiliation{VITCON Projectconsult GmbH, 07745 Jena, Germany}
\author{Michael L. W. Thewalt}
\affiliation{Department of Physics, Simon Fraser University, British Columbia V5A 1S6, Canada}
\author{Fay E. Hudson}
\affiliation{School of Electrical Engineering and Telecommunications, University of New South Wales, Sydney, NSW 2052, Australia}
\affiliation{Diraq Pty. Ltd., Sydney, NSW, Australia}
\author{Christopher C. Escott}
\affiliation{School of Electrical Engineering and Telecommunications, University of New South Wales, Sydney, NSW 2052, Australia}
\affiliation{Diraq Pty. Ltd., Sydney, NSW, Australia}
\author{Natalia Ares}
\affiliation{Department of Engineering Science, University of Oxford, Oxford OX1 3PH, United Kingdom}
\author{Stephen D. Bartlett}
\affiliation{Centre for Engineered Quantum Systems, School of Physics, University of Sydney, Sydney, New South Wales 2006, Australia}
\author{Andrea Morello}
\affiliation{School of Electrical Engineering and Telecommunications, University of New South Wales, Sydney, NSW 2052, Australia}
\author{Andre Saraiva}
\affiliation{School of Electrical Engineering and Telecommunications, University of New South Wales, Sydney, NSW 2052, Australia}
\affiliation{Diraq Pty. Ltd., Sydney, NSW, Australia}
\author{Arne Laucht} 
\affiliation{School of Electrical Engineering and Telecommunications, University of New South Wales, Sydney, NSW 2052, Australia}
\affiliation{Diraq Pty. Ltd., Sydney, NSW, Australia}
\author{Andrew S. Dzurak}
\email{a.dzurak@unsw.edu.au}
\affiliation{School of Electrical Engineering and Telecommunications, University of New South Wales, Sydney, NSW 2052, Australia}
\affiliation{Diraq Pty. Ltd., Sydney, NSW, Australia}
\author{Chih Hwan Yang}
\email{henry.yang@unsw.edu.au}
\affiliation{School of Electrical Engineering and Telecommunications, University of New South Wales, Sydney, NSW 2052, Australia}
\affiliation{Diraq Pty. Ltd., Sydney, NSW, Australia}

\date{\today}

\begin{abstract}

\textbf{
The encoding of qubits in semiconductor spin carriers has been recognised as a promising approach to a commercial quantum computer that can be lithographically produced and integrated at scale~\cite{zwanenburg2013silicon,veldhorst2014addressable,vandersypen2017interfacing,seedhouse2021quantum,zwerver2022qubits,vahapoglu2021single,hansen2022implementation,philips2022universal,vahapoglu2022coherent,weinstein2023universal}. 
However, the operation of the large number of qubits required for advantageous quantum applications~\cite{lekitsch2017blueprint,campbell2017roads,preskill2018quantum} will produce a thermal load exceeding the available cooling power of cryostats at millikelvin temperatures. 
As the scale-up accelerates, it becomes imperative to establish fault-tolerant operation above 1 kelvin, where the cooling power is orders of magnitude higher~\cite{almudever2017engineering,petit2018spin,yang2020operation,petit2020universal,camenzind2022hole,petit2022design}. 
Here, we tune up and operate spin qubits in silicon above 1 kelvin, with fidelities in the range required for fault-tolerant operation at such temperatures~\cite{raussendorf2007fault,wang2011surface,fowler2012surface}. 
We design an algorithmic initialisation protocol to prepare a pure two-qubit state even when the thermal energy is substantially above the qubit energies, and incorporate radio-frequency readout to achieve fidelities up to 99.34 per cent for both readout and initialisation.
Importantly, we demonstrate a single-qubit Clifford gate fidelity of 99.85 per cent, and a two-qubit gate fidelity of 98.92 per cent. 
These advances overcome the fundamental limitation that the thermal energy must be well below the qubit energies for high-fidelity operation to be possible, surmounting a major obstacle in the pathway to scalable and fault-tolerant quantum computation.
}

\end{abstract}

\maketitle

Quantum computation is an emerging technology that promises to outperform classical transistor-based computation in certain tasks and solve currently intractable problems. To realise its promised benefits, we require large arrays of qubits that can operate within densely packed cryogenic platforms, and this may necessitate operating at temperatures well above the millikelvin regime~\cite{campbell2017roads,almudever2017engineering,petit2018spin,yang2020operation,petit2020universal,camenzind2022hole,petit2022design}. Spins  in semiconductor quantum dot arrays have been considered a rising candidate for this undertaking, thanks to their low error rates, long information hold time, and industrial compatibility~\cite{veldhorst2014addressable,vandersypen2017interfacing,gonzalez-zalba2021scaling,zwerver2022qubits}. Previous experiments were primarily undertaken in millikelvin environments, where thermal noise is minimised, for studying intrinsic qubit properties and developing operation techniques. Initial studies of spin qubit operation above $\SI{1}{\kelvin}$ have verified the feasibility, despite suffering from degraded state-preparation-and-measurement (SPAM) and gate fidelities~\cite{petit2018spin,yang2020operation,petit2020universal,camenzind2022hole}. These challenges can be tackled by combining new device designs and novel engineering techniques, in areas from initialisation to control and readout. In this work, we operate electron spin qubits in silicon with SPAM and universal logic fidelities approaching the quantum error correction (QEC) standards based on surface codes~\cite{raussendorf2007fault,wang2011surface,fowler2012surface}, which has only been demonstrated with nuclear spins at millikelvin temperatures so far~\cite{madzik2022precision}. We enable deterministic two-qubit initialisation in silicon above $\SI{1}{\kelvin}$ via an entropy-transferring algorithmic initialisation protocol based on two-qubit logic and single-shot readout. This capability is pivotal in areas such as quantum information processing, quantum memories, and analogue quantum simulations. The excellent high-temperature performance of semiconductor spin qubits underpins their potential to act as the building block of large-scale quantum processors with integrated classical control electronics. Additionally, we present a complete error analysis in the two-qubit space, and characterise every aspect of operation at different temperatures and external magnetic field to open up further studies.

\begin{figure*}[ht!]
    \includegraphics[angle = 0]{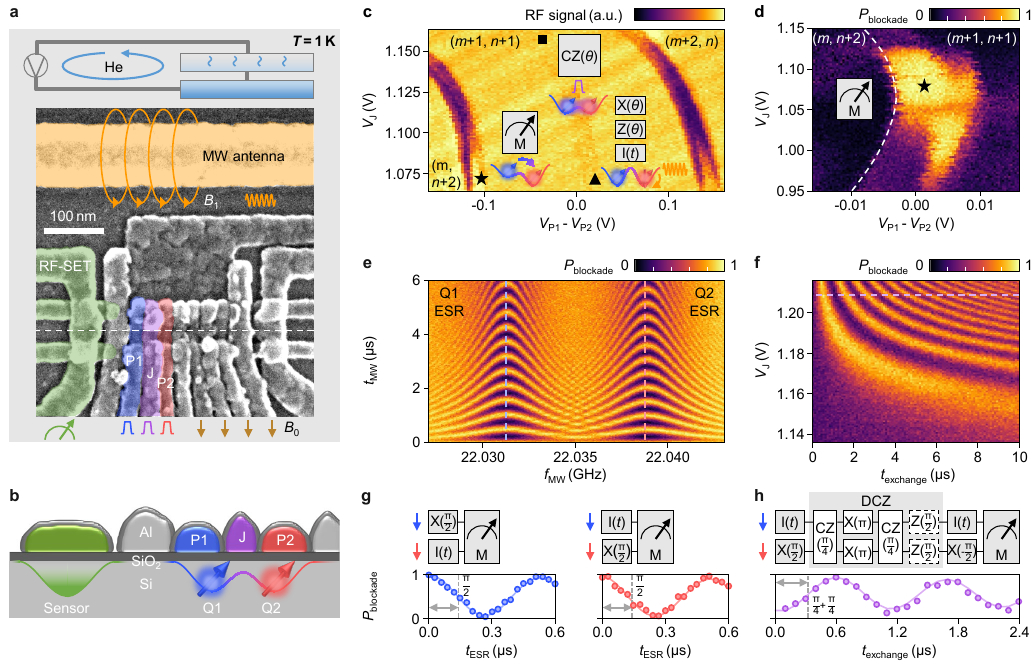}
    \caption{\textbf{Device and basic operation. ${\mathbf P_\mathrm{blockade}}$ is unscaled in all data.} 
    \textbf{a}, Schematic experimental setup, with a scanning electron micrograph of a device nominally identical to that used in this work. Active gate electrodes and the microwave antenna are highlighted with colours. An external DC magnetic field $B_0$ and the antenna-generated AC magnetic field $B_1$ are indicated with arrows. The system operates at $T = \SI{1}{\kelvin}$, unless otherwise specified.
    \textbf{b}, Simulated device cross-section view, and a schematic indicating the intended quantum dots, the electron spin qubits, and the RFSET sensor.
    \textbf{c}, Charge stability diagram as a function of P1, P2 voltage detuning and the J gate voltage $V_\mathrm{J}$, showing the operation regime. The operation points for readout (M), single-qubit (X, Z, I) and two-qubit controlled phase (CZ) operation are labelled as star (\scalebox{1.1}{$\star$}), triangle (\scalebox{0.75}{$\blacktriangle$}) and square (\scalebox{0.5}{$\blacksquare$}), respectively. The insets schematically show the operations that are performed at each position.
    \textbf{d}, Probability of detecting a blockaded state, $P_\mathrm{blockade}$, when preparing $\ket{\downarrow\downarrow}$ and reading out at different $V_\mathrm{J}$ and P1, P2 voltage detuning. The readout location for subsequent experiments is set amid the readout window which appears as the high-$P_\mathrm{blockade}$ region.
    \textbf{e}, Single-qubit Rabi oscillations at $V_\mathrm{J}=\SI{1.1}{\volt}$ as a function of microwave frequency $f_\mathrm{MW}$ and pulse time $t_\mathrm{MW}$.
    \textbf{f}, Decoupled controlled phase ($\mathrm{DCZ}$) oscillations as a function of exchange time $t_\mathrm{exchange}$ and $V_\mathrm{J}$.
    \textbf{g}, Calibration of the single-qubit $\mathrm{X}(\pi/2)$ gates.
    \textbf{h}, Calibration of the two-qubit DCZ gate.
    }
    \label{fig:main_fig_1}
\end{figure*}

\subsection{Device and two-qubit operation}

We conduct our study on a prototype two-qubit processor based upon a silicon-metal-oxide-semiconductor (SiMOS) double quantum dot (Fig.~\ref{fig:main_fig_1}~a--b). Each qubit is encoded in the spin state of an unpaired electron~\cite{veldhorst2015spin,leon2020coherent}. The device incorporates multi-level aluminium gate-stacks~\cite{angus2008silicon,lim2009observation} fabricated on an isotopically enriched ${}^{28}$Si substrate with $\SI{50}{ppm}$ residual ${}^{29}$Si~\cite{becker2010enrichment}. The quantum dots are electrostatically defined in areas of around $\SI{80}{\nano\meter^2}$ underneath the plunger gates (P1, P2), at the Si/SiO$_2$ interface. An exchange gate (J) controls the inter-dot separation and two-qubit exchange~\cite{loss1998quantum,petta2005coherent,cifuentes2023bounds} at an exponential rate of $\SI{20}{dec/V}$. A radio-frequency single-electron transistor (RFSET)~\cite{angus2008silicon} operating at $\SI{0.21}{\giga\hertz}$ is employed for single-shot charge readout, with a nominal signal integration time $t_\mathrm{integration}=\SI{50}{\micro\second}$. See Methods~\ref{methods:measurement_setup} for a description of the complete setup.

To form the qubits (labelled Q1 and Q2), we load an odd number of electrons in the P1 and P2 dots (Fig.~\ref{fig:main_fig_1}~c). We operate in the two-qubit basis of $\{\ket{\downarrow\downarrow}, \ket{\downarrow\uparrow}, \ket{\uparrow\downarrow}, \ket{\uparrow\uparrow}\}$, where $\downarrow$ and $\uparrow$ denote spin-down and spin-up, and measure the states using parity readout, a type of qubit readout based on Pauli spin blockade (PSB)~\cite{ono2002current,lai2011pauli,seedhouse2021pauli}. In the PSB region, located at the inter-dot charge transition point (Fig.~\ref{fig:main_fig_1}~d), charge movement is blockaded when the two qubits are parallel, i.e., $\ket{\downarrow\downarrow}$ and $\ket{\uparrow\uparrow}$. We prefer to load three electrons in at least one of the dots to avoid the small excitation energy -- and consequently narrow PSB range -- caused by the valley excitation in the case of a single electron. After locating the PSB region, algorithmic initialisation is employed to deterministically prepare a two-qubit state. Single-qubit gates are based on microwave pulses at the electron-spin resonant frequencies ($f_\mathrm{ESR}$) delivered through the antenna, combined with phase rotations, and two-qubit gates take the form of decoupled controlled phase gate ($\mathrm{DCZ}$)~\cite{watson2018programmable,xue2022quantum} (Fig.~\ref{fig:main_fig_1}~g--h). See Methods~\ref{methods:device_tuneup} for the tune-up details.

Fig.~\ref{fig:main_fig_1}~e--f show the coherent and stable single-qubit Rabi and two-qubit exchange oscillations, taken at $T = \SI{1}{\kelvin}$ and external magnetic field $B_0 = \SI{0.79}{\tesla}$. Benefiting from the low charge and spin noise level (Extended Data Fig.~\ref{fig:fESR_noise}~b--d), feedback on the gate voltage levels, spin resonance frequencies, and microwave amplitudes~\cite{philips2022universal} are not used, which reduces the number of feedback parameters by 7 and lowers the time and computation cost. Feedback on the RFSET sensor is retained to automatically maintain the readout signal level over long periods of time~\cite{yang2011dynamically}.

\begin{figure*}[ht!]
    \includegraphics[angle = 0]{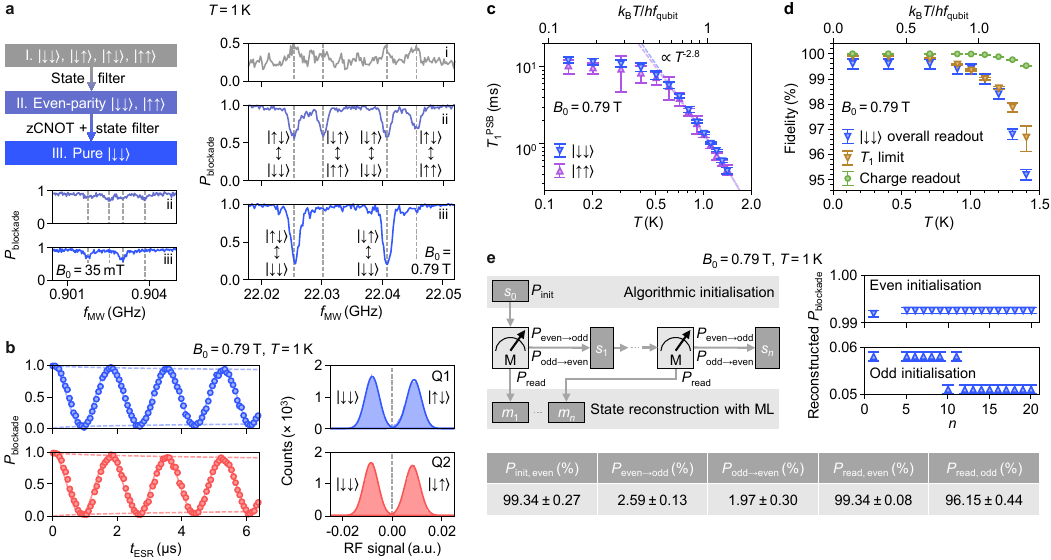}
    \caption{\textbf{Initialisation and readout.} 
    \textbf{a}, Two-qubit algorithmic initialisation and the outcomes at $B_0 = \SI{0.79}{\tesla}$ and $\SI{35}{\milli\tesla}$, both at $T = \SI{1}{\kelvin}$. i represents the conventional ramped initialisation over a duration of $\SI{100}{\micro\second}$, and ii, iii are the partial and full algorithmic initialisation. The traces are taken at $V_\mathrm{J}=\SI{1.2}{\volt}$ where exchange is on, with dashed lines indicating the locations of the four state transitions.
    \textbf{b}, Left: resonantly driven Rabi oscillations of individual qubits for a short $t_\mathrm{MW}$ and averaged over 500 shots, at $B_0 = \SI{0.79}{\tesla}$ and $T = \SI{1}{\kelvin}$. $P_\mathrm{blockade}$ is unscaled in both traces. Right: corresponding charge readout histograms. The signal integration time $t_\mathrm{integration}$ is $\SI{50}{\micro\second}$.
    \textbf{c}, PSB relaxation time $T_1^\mathrm{PSB}$ for $\ket{\downarrow\downarrow}$ and $\ket{\uparrow\uparrow}$ as a function of temperature from $\SI{0.14}{\kelvin}$ to $\SI{1.4}{\kelvin}$, at $B_0 = \SI{0.79}{\tesla}$.
    \textbf{d}, Measured readout fidelity and estimated $T_1$-limited spin readout fidelity of $\ket{\downarrow\downarrow}$, and charge readout fidelity as a function of temperature from $\SI{0.14}{\kelvin}$ to $\SI{1.4}{\kelvin}$, at $B_0 = \SI{0.79}{\tesla}$ with $t_\mathrm{integration}=\SI{50}{\micro\second}$.
    \textbf{e}, State reconstruction and state-preparation-and-measurement (SPAM) error analysis using repeated PSB readout at $B_0 = \SI{0.79}{\tesla}$ and $T = \SI{1}{\kelvin}$. We initialise $s_0=\ket{\downarrow\downarrow}$ using the algorithmic initialisation or $\ket{\uparrow\downarrow}$ by $\pi$-pulsing on Q1 after the algorithmic $\ket{\downarrow\downarrow}$ initialisation. We then perform $n$ PSB readouts, through which the state evolves into $s_n$. Finally, we apply machine learning on the readout outcomes $m_1$--$m_n$ to extract the initialisation, readout and spin-flip probabilities, and reconstruct the states. The results are shown in the table and the plots.
    }
    \label{fig:main_fig_2}
\end{figure*}

\subsection{Initialisation and readout}

Initialisation is a prerequisite to all qubit operations, and SPAM is as important to quantum computation as universal logic. At millikelvin temperatures, where the qubit energy $hf_\mathrm{qubit}$ is much greater than the thermal energy $k_\mathrm{B}T$, electron spin qubit initialisation may rely on intrinsic polarisation mechanisms such as spin-dependent tunneling from a spin carrier reservoir~\cite{elzerman2004single,morello2010single,mills2022high}, PSB~\cite{fogarty2018integrated,zhao2019single,petit2020universal,camenzind2022hole} or additionally accelerated relaxation~\cite{yang2013spin,yang2020operation,blumoff2022fast}. Higher-fidelity single-qubit state preparation can be achieved using initialisation by measurement~\cite{yoneda2020quantum,johnson2022beating}, and conditional single-qubit pulses~\cite{philips2022universal,kobayashi2023feedback}. These approaches either partially rely on intrinsic polarisation, or require readout with a reservoir, which are incompatible with operation at elevated temperatures. In this work, we design a generic two-qubit algorithmic initialisation protocol that works in conditions where $hf_\mathrm{qubit}$ is comparable to or less than $k_\mathrm{B}T$. The method is applicable to a large-scale qubit array, where initialisation and readout are performed pairwise~\cite{fogarty2018integrated,zhao2019single,yang2020operation}.

Fig.~\ref{fig:main_fig_2}~a shows the algorithmic initialisation protocol, with the resulting two-qubit state at each stage at $T = \SI{1}{\kelvin}$ and $B_0 = \SI{0.79}{\tesla}$. The state composition can also be verified from the ESR spectrum when exchange is on. Stage I, the outcome of a $\SI{100}{\micro\second}$ ramp into the operation point, has a mixture of $\ket{\downarrow\downarrow}$, $\ket{\downarrow\uparrow}$, $\ket{\uparrow\downarrow}$ and $\ket{\uparrow\uparrow}$ states, and the measured ESR transitions are almost indistinguishable. To purify the state hereafter, we move to Stage II and perform the ramped initialisation followed by parity readout. The parity readout preserves the even-parity states as long as it is performed faster than the spin relaxation time~\cite{seedhouse2021pauli}. The algorithm re-enters Stage II if the readout is unblockaded, and proceeds otherwise. Here the output is a mixture of $\ket{\downarrow\downarrow}$ and $\ket{\uparrow\uparrow}$ with the odd-parity states filtered out, which can be identified from the associated ESR transitions. For further purification, we move to Stage III with a zero controlled NOT gate (zCNOT)~\cite{madzik2022precision} to convert the remnant $\ket{\uparrow\uparrow}$ into $\ket{\uparrow\downarrow}$, which is then filtered out due to the odd parity. The ESR measurement after this stage shows two predominant transitions pertaining to $\ket{\downarrow\downarrow}$, with amplitude limited by the two-qubit exchange ~\cite{huang2019fidelity}. From these spectra, we extract an initialisation fidelity of $\SI{99.6}{\percent}$. See Methods~\ref{methods:algorithmic_initialisation} for the protocol details and Supplementary Fig.~\ref{fig:downdown_amplitude_extraction} for the ESR spectra analysis.

This initialisation protocol is robust to low $B_0$, and we expect it to be limited by the fidelities of control and readout, on which the protocol relies, and their time scale relative to that of spin relaxation and thermalisation. Fig.~\ref{fig:main_fig_2}~a shows the initialisation outcomes at $B_0 = \SI{35}{\milli\tesla}$, where $k_\mathrm{B}T$ is more than 20 times larger than $hf_\mathrm{qubit}$. The initialisation fidelity remains above $\SI{99}{\percent}$ at $B_0=\SI{85}{\milli\tesla}$ and above $\SI{90}{\percent}$ at $B_0=\SI{35}{\milli\tesla}$, but the ESR amplitude is further reduced due to the deviation from parity readout with the small $dE_\mathrm{Z}$ (also see Extended Data Fig.~\ref{fig:B0_dependence}~f). In most operating conditions, the protocol takes around 3 iterations, and the initialisation process spans between 100 and $\SI{200}{\micro\second}$ (see Extended Data Fig.~\ref{fig:algorithmic_init}~c--d).

When addressing individual qubits without pulsing exchange, we obtain Rabi oscillations with a raw amplitude of $0.950$--$0.966$ for the two qubits (Fig.~\ref{fig:main_fig_2}~b) at $T = \SI{1}{\kelvin}$ and $B_0 = \SI{0.79}{\tesla}$. We see that both Rabi oscillations start from $0.996$ and go down to $0.030$--${0.046}$ after a $\pi$ duration. We thus estimate that the initialisation and the overall readout fidelities are $\SI{99.6}{\percent}$ for $\ket{\downarrow\downarrow}$, and at least $\SI{95.0}{\percent}$ for $\ket{\downarrow\uparrow}$ and $\ket{\uparrow\downarrow}$.

At $T = \SI{1}{\kelvin}$ and $B_0 = \SI{0.79}{\tesla}$, the relaxation time $T_1$, which is the time for a single spin flip at the single-qubit operation point, is $12.23\pm\SI{2.11}{\milli\second}$ and the PSB relaxation time $T_1^\mathrm{PSB}$, which is the life time of a blockaded state at the PSB region, is $1.36\pm\SI{0.06}{\milli\second}$. We use $t_\mathrm{integration}=\SI{50}{\micro\second}$, a time much shorter than $T_1^\mathrm{PSB}$, to achieve a charge readout fidelity of $\SI{99.7}{\percent}$. With these considered, the Rabi amplitude is most likely limited by control errors and the diabaticity in reading out odd-parity states.



Fig.~\ref{fig:main_fig_2}~c and d show the temperature dependence of these metrics between $\SI{0.14}{\kelvin}$ and $\SI{1.4}{\kelvin}$ at $B_0 = \SI{0.79}{\tesla}$. The PSB relaxation times scale with $T^{-2.8}$ above $\SI{0.5}{\kelvin}$, dropping by ten fold to $\SI{0.45}{\milli\second}$ at $T=\SI{1.4}{\kelvin}$. This reduction implies that future readout techniques should avoid compromising on the total readout time. Below $\SI{1}{\kelvin}$, the overall readout fidelity for $\ket{\downarrow\downarrow}$ falls slowly and appears to be limited by neither $T_1^\mathrm{PSB}$ or charge readout, whereas above $\SI{1}{\kelvin}$, these two limitations are present. $T_1$-induced errors increase at a higher rate than charge readout errors and appear as the dominating factor towards even higher temperatures. Overall, SPAM around $\SI{1}{\kelvin}$ is comparable to that at millikelvin temperatures and remains workable at least until $\SI{1.4}{\kelvin}$.

Lastly, we test repeated parity readout at $T=\SI{1}{\kelvin}$. We apply machine learning to infer the parity readout errors and probabilities during SPAM, and reconstruct the true initial state parity using the cumulative readout outcomes~\cite{yoneda2020quantum}. In principle, not only are the even-parity states protected by PSB, but the odd-parity states are also likely to remain odd-parity after pulsing quickly back from the readout. Fig.~\ref{fig:main_fig_2}~e shows this protocol along with the results of our analysis. See Methods~\ref{methods:SPAM_analysis} for details on the machine learning approach. The state preparation and measurement fidelities are captured by $P_\mathrm{init}$ and $P_\mathrm{read}$, and the probabilities of state changes during each readout cycle are captured by $P_\mathrm{even\rightarrow odd}$ and $P_\mathrm{odd\rightarrow even}$. With the algorithmic $\ket{\downarrow\downarrow}$ initialisation and using 20 readout cycles ($n=20$), we infer $P_\mathrm{init, even}$, $P_\mathrm{init, odd}$ to be $99.34\pm\SI{0.27}{\percent}$, $94.67\pm\SI{0.73}{\percent}$, and $P_\mathrm{read, even}$, $P_\mathrm{read, odd}$ to be $99.34\pm\SI{0.08}{\percent}$, $96.15\pm\SI{0.44}{\percent}$ respectively. For $\ket{\downarrow\downarrow}$ initialisation with $n=5$, the reconstructed $P_\mathrm{blockade}$ increases from $\SI{99.2}{\percent}$ to $\SI{99.3}{\percent}$, and for $\ket{\uparrow\downarrow}$ initialisation with $n=12$, $P_\mathrm{blockade}$ decreases from $\SI{5.8}{\percent}$ to $\SI{5.1}{\percent}$. The full set of probabilities are detailed in Supplementary Fig.~\ref{fig:SPAM_analysis}.

\begin{figure*}[ht!]
    \includegraphics[angle = 0]{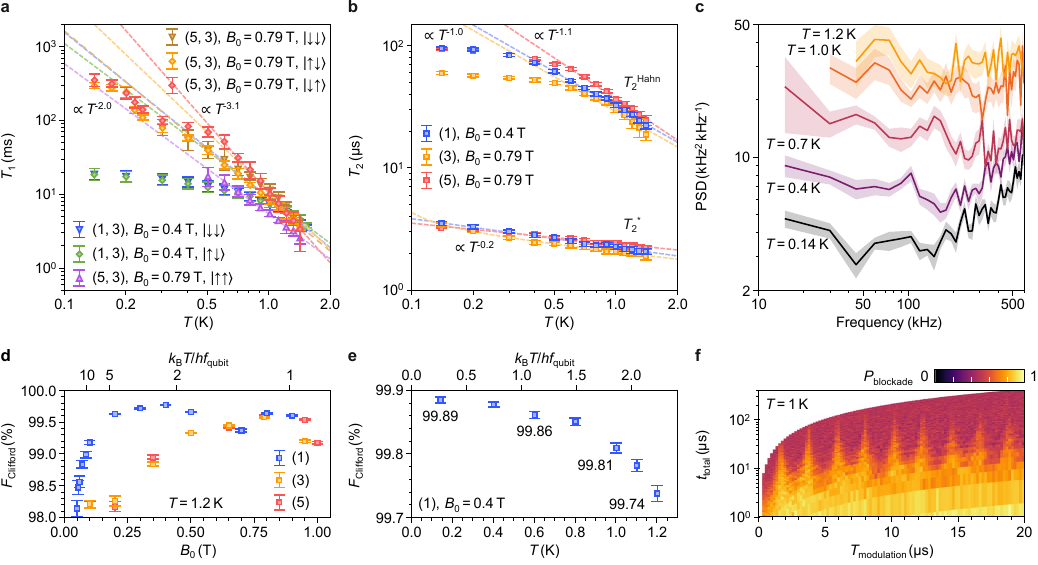}
    \caption{\textbf{Single-qubit performance.} 
    \textbf{a}, Relaxation time $T_1$ of different states as a function of temperature from $\SI{0.14}{\kelvin}$ to $\SI{1.4}{\kelvin}$, in different charge configurations for Q1, Q2.
    \textbf{b}, Dephasing times, $T_2^{*}$ and $T_2^\mathrm{Hahn}$, as a function of temperature from $\SI{0.14}{\kelvin}$ to $\SI{1.4}{\kelvin}$, in different charge configurations for Q1.
    \textbf{c}, Single-qubit noise power spectral density (PSD) of Q1 based on the Carr-Purcell-Meiboom-Gill (CPMG) protocol~\cite{cywinski2008enhance,alvarez2011measuring,medford2012scaling} at different temperatures at $B_0=\SI{0.79}{\tesla}$.
    \textbf{d}, Single-qubit Clifford gate fidelity as a function of $B_0$ from $\SI{50}{\milli\tesla}$ to $\SI{1}{\tesla}$ at $T = \SI{1.2}{\kelvin}$, in different charge configurations for Q1.
    \textbf{e}, Single-qubit Clifford gate fidelity as a function of temperature at $B_0 = \SI{0.4}{\tesla}$, with the one-electron configuration.
    \textbf{f}, Demonstration of a dressing protocol, the SMART protocol on Q1~\cite{hansen2022implementation} at $B_0 = \SI{0.5}{T}$ and $T = \SI{1}{\kelvin}$ with $f_\mathrm{Rabi}=\SI{1.44}{\mega\hertz}$, showing periodical modulation optima.
    }
    \label{fig:main_fig_3}
\end{figure*}

\begin{figure*}[ht!]
    \includegraphics[angle = 0]{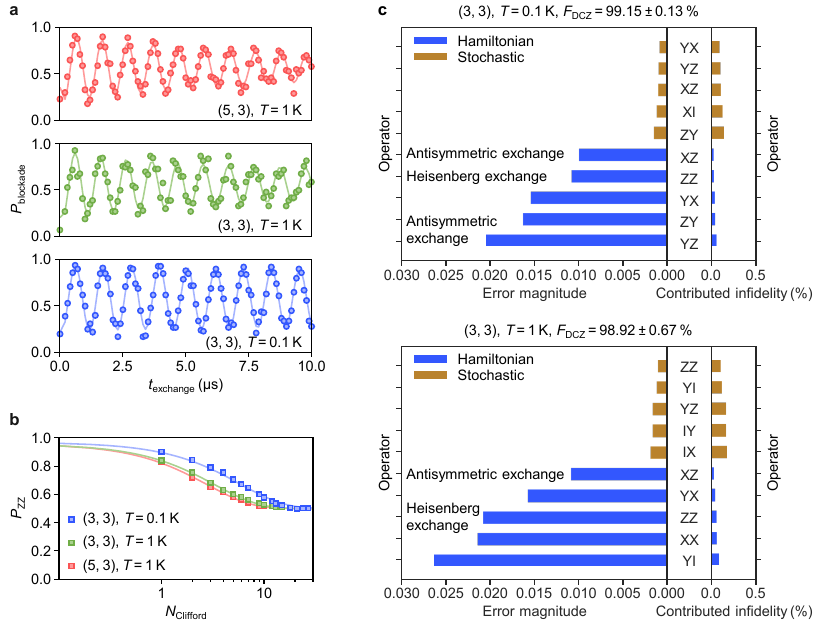}
    \caption{\textbf{Two-qubit performance.} 
    \textbf{a}, $\mathrm{DCZ}$ oscillations at $B_0=\SI{0.79}{\tesla}$, $T = \SI{0.1}{\kelvin}$ and $\SI{1}{\kelvin}$.
    \textbf{b}, Two-qubit randomised benchmarking at $B_0=\SI{0.79}{\tesla}$, $T = \SI{0.1}{\kelvin}$ and $\SI{1}{\kelvin}$.
    \textbf{c}, Breakdown of error channels using pyGSTi~\cite{nielsen2022pyGSTio}, based on the final FBT error generators at $B_0=\SI{0.79}{\tesla}$, $T = \SI{0.1}{\kelvin}$ and $\SI{1}{\kelvin}$, with the error magnitudes plotted towards the left and their contributed infidelities plotted towards the right. Each sub-figure includes the five largest contributing channels for both Hamiltonian (blue) and Stochastic (gold) errors respectively. We note that Hamiltonian errors contributes to the infidelity in second order, but Stochastic errors contribute in first order~\cite{blume-kohout2022a}. Common error channels are labeled with their physical interpretations.
    }
    \label{fig:main_fig_4}
\end{figure*}

\subsection{Single-qubit performance}

Relaxation time ($T_\mathrm{1}$) and dephasing time ($T_\mathrm{2}$), as well as the single-qubit control fidelities tend to be excellent in silicon~\cite{yoneda2018quantumdot,yang2019silicon,muhonen2014storing}, with single-qubit fidelity moderately above $\SI{99}{\percent}$ previously attained at $T = \SI{1.1}{\kelvin}$~\cite{petit2020universal}. In this device, we expect the reduced charge and magnetic noise, as well as the strong voltage confinement to extend $T_2$. Moreover, fast microwave driving is applied to maximise the number of quantum gates before the qubit fully decays.

We first study $T_\mathrm{1}$ and $T_\mathrm{2}$ (Fig.~\ref{fig:main_fig_3}~a--b) in this device in the $(1,3)$ and $(5,3)$ charge states near the optimal $B_0$. The dominating noise sources -- charge noise, Johnson noise, Orbach and Raman phonon scattering -- may change at different temperature ranges, giving rise to an intricate temperature dependence of $T_1$~\cite{petit2018spin}, and the involvement of both qubits in parity readout adds to the complication. Nevertheless, we note that the temperature dependence of all states fall between $T^{-2.0}$ and $T^{-3.1}$ above $\SI{1}{\kelvin}$. Additionally, the thermal equilibrium shifts from $\ket{\downarrow\downarrow}$ when $k_\mathrm{B}T\ll hf_\mathrm{qubit}$ to a mixed state when $k_\mathrm{B}T\geq hf_\mathrm{qubit}$. See Extended Data Fig.~\ref{fig:T1s_raw} for a more detailed $T_\mathrm{1}$ analysis. The temperature dependence of $T_2^\mathrm{Hahn}$ in different configurations fall between $T^{-1}$ and $T^{-1.1}$, while $T_2^*$ scale uniformly to $T^{-0.2}$. The temperature scaling power of both $T_\mathrm{1}$ and $T_\mathrm{2}$ are lower than those in most of the previous results~\cite{petit2018spin,yang2020operation,camenzind2022hole}. The bias of Z errors (dephasing noise) to X errors (depolarisation noise) can be indicated by the $T_1/T_2$ ratio, which is shown in Extended Data Fig.~\ref{fig:fESR_noise}~a.

We perform noise spectroscopy based on the Carr-Purcell-Meiboom-Gill (CPMG) protocol~\cite{cywinski2008enhance,alvarez2011measuring,medford2012scaling}, which uses a single qubit as a noise probe to detect the noise power spectral density (PSD) at different temperatures. As shown in Fig.~\ref{fig:main_fig_3}~c, the overall noise level rises with temperature within the detectable frequency range. In the white noise regime ($\geq\SI{200}{\kilo\hertz}$), the noise power spectral density increases by an order of magnitude from $\SI{0.14}{\kelvin}$ to $\SI{1.2}{\kelvin}$. We notice an increase in the apparent PSD at higher frequencies that we characterise as blue noise, possibly due to accumulated microwave pulse miscalibration, or an effect from the high-power driving~\cite{philips2022universal,undseth2023hotter}. See Extended Data Fig.~\ref{fig:fESR_noise}~e for the full set of power noise spectral density traces, and Extended Data Fig.~\ref{fig:fESR_noise}~f for another measurement of the microwave effect.

With the strong capability of noise decoupling ($T_2^\mathrm{Hahn}/T_2^{*}$ reaching 10) and fast spin rotations at $T = \SI{1}{\kelvin}$, the optimised single-qubit Clifford fidelity in randomised benchmarking (RB)~\cite{knill2008randomized,magesan2011scalable,magesan2012efficient} is up to $99.85\pm \SI{0.01}{\percent}$ (see Supplementary Fig.~\ref{fig:long_RB_dual_projection} for raw data). Correction of crosstalk is crucial due to the relatively small difference in $f_\mathrm{ESR}$. See Methods~\ref{methods:crosstalk} for crosstalk correction and Methods~\ref{methods:RB} for the implementation of RB. As shown in Fig.~\ref{fig:main_fig_3}~d, we observe a fidelity reduction towards low $B_0$, mostly governed by the maximum Rabi frequency due to crosstalk (Extended Data Fig.~\ref{fig:crosstalk}~c), or near an excited state degeneracy at high $B_0$ where the  dephasing is enhanced. The one-electron qubit operates with $99.18\pm\SI{0.03}{\percent}$ at $B_0 = \SI{0.1}{\tesla}$, where $f_\mathrm{qubit}=\SI{2.8}{GHz}$ and $k_\mathrm{B}T=7hf_\mathrm{qubit}$. The qubits are operable at $B_0$ as low as $\SI{25}{\milli\tesla}$. See Extended Data Fig.~\ref{fig:B0_dependence} for the full study. Such results suggest the possibility of ultra-low $B_0$ operation to significantly reduce the hardware and power cost~\cite{zhao2019single}. Going from $\SI{0.14}{\kelvin}$ to $\SI{1}{\kelvin}$, we see a reduction of less than $\SI{1}{\percent}$ in the Clifford gate fidelity (Fig.~\ref{fig:main_fig_3}~e). With continuous driving in the single-qubit RB, we expect the incoherent errors to mainly come from the white noise floor, and the blue noise from microwave pulsing.

Furthermore, we extend the recent demonstration of a dressing protocol, the SMART protocol for a single qubit from $\SI{0.1}{\kelvin}$~\cite{seedhouse2021quantum,hansen2021pulse,hansen2022implementation} to $\SI{1}{\kelvin}$. See Extended Data Fig.~\ref{fig:crosstalk}~e for the gate sequence. We observe extended coherence at certain periods of microwave modulation $T_\mathrm{modulation}$, that corresponds to the optimal noise cancellation conditions for the modulation. This demonstration substantiates the potential to continuously drive a large number of spin qubits with a global field above $\SI{1}{\kelvin}$ in future architectures.

\subsection{Two-qubit performance}

Two-qubit gate fidelity in silicon has recently seen agreeable improvement, reaching the fault-tolerant requirements~\cite{noiri2022fast,xue2022quantum,madzik2022precision,mills2022twoqubit,tanttu2023stability}, and extending this to above $\SI{1}{\kelvin}$ becomes of great interest. We incorporate a decoupling X($\pi$) gate on individual qubits in the middle of the $\mathrm{CZ}$ gate to extend coherence and cancel Stark shift-induced phase errors~\cite{watson2018programmable,xue2022quantum}. This constitutes the decoupled controlled phase (DCZ) operation. With approximately the same level of exchange, the quality factor of the $\mathrm{DCZ}$ oscillation is well above 100 at $T=\SI{0.1}{\kelvin}$ and remains at least 50 above $\SI{1}{\kelvin}$, exhibiting a coherence reduction similar to that in $T_2^\mathrm{Hahn}$ (Fig.~\ref{fig:main_fig_4}~a). See Extended Data Fig.~\ref{fig:DCZ_raw} for the full characterisation of exchange.

We assess the $\mathrm{DCZ}$ gate metrics using two-qubit RB and fast Bayesian tomography (FBT)~\cite{evans2022fast,su2023characterizing} (see Methods~\ref{methods:RB} and Methods~\ref{methods:FBT}). From the raw RB fidelities, we first obtain the interleaved-RB (IRB) ratio $F_\mathrm{interleaved}/F_\mathrm{reference}=99.8\pm\SI{0.2}{\percent}$ at $T=\SI{0.1}{\kelvin}$, and $97.7\pm\SI{1.5}{\percent}$ at $T=\SI{1}{\kelvin}$. Instead of the true DCZ fidelity, this ratio mainly reflects the combined effect of dephasing during $t_\mathrm{exchange}$ and echoing in the DCZ gate, and the results can be understood from the stronger temperature-dependence of $T_2^\mathrm{Hahn}$ compared to that of $T_2^\mathrm{*}$. We also note the numerical instabilities in the IRB ratio, which results in large error bars. In comparison, FBT extracts DCZ fidelities of $99.15\pm\SI{0.13}{\percent}$ at $T=\SI{0.1}{\kelvin}$, and $98.92\pm\SI{0.67}{\percent}$ at $T=\SI{1}{\kelvin}$. Here, a single-qubit gate on one qubit always leaves the other qubit idling, which considerably limits the single-qubit process fidelities (see Supplementary Fig.~\ref{fig:error_generators}) and consequently the Clifford fidelity in two-qubit RB, even at $T=\SI{0.1}{\kelvin}$ (Fig.~\ref{fig:main_fig_4}~b). However, the reduction in the Clifford fidelity from $\SI{0.1}{\kelvin}$ to $\SI{1}{\kelvin}$ mainly originates from the degradation of the $\mathrm{DCZ}$ gate, exhibiting a similar factor.

When examining the fidelity results, we are also interested in understanding the dominant error sources behind the $\mathrm{DCZ}$ gate infidelity, and their variation at different temperatures. FBT is a flexible and efficient gate set process tomography, that allows us to extract gate errors from randomised sequence runs~\cite{evans2022fast,su2023characterizing}.  To categorise the gate errors, we post process the tomography results obtained by FBT using tools for decomposing errors implemented in the pyGSTi package~\cite{nielsen2022pyGSTio, blume-kohout2022a}. We first convert the error process matrices into error generators, which are then decomposed into four error generator subspaces, the Hamiltonian, stochastic, correlated stochastic, and active error generator (Supplementary Fig.~\ref{fig:error_generators} and Fig.~\ref{fig:error_channels}). See Methods~\ref{methods:FBT} and Methods~\ref{methods:error_taxonomy} for details. We display the five largest components from Hamiltonian error and stochastic error in Fig.~\ref{fig:main_fig_4}~c. We see a change in the error landscape when we go from $\SI{0.1}{\kelvin}$ to $\SI{1}{\kelvin}$. At both $\SI{0.1}{\kelvin}$ and $\SI{1}{\kelvin}$, exchange noise presents itself as one of the main noise sources. We observe terms represented by the Heisenberg exchange, and the antisymmetric exchange, also known as Dzyaloshinskii–Moriya (DM) interaction. We expect that the antisymmetric exchange leads to Hamiltonian error terms that couple in a ZY- or ZX-like manner. While this is an important source of error that should be reduced, we note that the main contributions to the infidelity itself come in the form of stochastic errors, which contribute linearly to the infidelity.

We identify that there is not a single stochastic error source that dominates the DCZ gate infidelity (unlike the single qubit gates as shown in Supplementary Fig.~\ref{fig:error_channels}). This means that it will be non-trivial to reduce the stochastic error significantly. We attribute some of these errors to the slight differences in coupling strength each time a DCZ is executed. It would require advanced pulse shaping techniques to eliminate the non-Markovian noise sources causing these inconsistencies.

We also perceive an asymmetry in the error types: the errors on Q1 being very different from those on Q2. Despite the symmetric operation point~\cite{noiri2022fast,xue2022quantum,reed2016reduced,martins2016noise} in the $(3,3)$ charge state, the error symmetry may be convoluted by the local environmental factors of individual qubits, such as $\mathrm{Si/SiO_2}$ interface roughness and charge noise~\cite{cifuentes2023bounds}. If we could controllably choose the asymmetric operation in a particular dot direction, it would be potentially beneficial to QEC with tailored surface codes.

The detailed nature of the dominant error processes in silicon spin qubits offers significant opportunities for innovations in codes and architectures. We observe a bias in the error rates towards dephasing, generally larger than $100:1$ up to at least $T=\SI{1.5}{\kelvin}$, for which increased fault-tolerant thresholds are possible~\cite{tuckett2018ultrahigh,tuckett2019tailoring,tuckett2020fault}. To exploit such gains, further research would be needed to characterise the process of error syndrome extraction, where each cycle involves SPAM on the ancilla qubits during which the data qubits can undergo decoupling. We expect a moderate decrease in the noise bias from decoupling with increasing temperatures (see Extended Data Fig.~\ref{fig:fESR_noise}~a), but this may not be true for even higher temperatures. The CZ-type operation we employ as a 2-qubit gate can be bias-preserving, but fully exploiting this bias for QEC will require syndrome extraction circuit design to avoid injection of spin-flip errors from SPAM of the ancilla qubits into the data qubits.

\subsection{Outlook}
The use of algorithmic qubit initialisation and the realisation of high-fidelity universal logic in this work bring SiMOS spin qubits at temperatures above $\SI{1}{\kelvin}$ into the realm of fault tolerance. Furthermore, the proven ability to operate at low $B_\mathrm{0}$ will benefit large-scale global control~\cite{vahapoglu2021single,vahapoglu2022coherent} with low driving frequency, and reduce the cost of microwave instrumentation. This further strengthens semiconductor spin qubits as an affordable approach. Aside from setting the benchmark for initialisation, control and readout fidelities at elevated temperatures, we present here a complete study of the properties of the two-qubit system. We show certain robustness against the charge configuration and the applied magnetic field above $T=\SI{1}{\kelvin}$, which is important to large scale operation. The similar temperature dependence of $T_1$ and $T_2$ in different configurations above $T=\SI{1}{\kelvin}$ suggests a potentially weaker effect from qubit variability~\cite{cifuentes2023bounds} at such temperatures.

The techniques explored in this work, including the analysis methods, machine learning for SPAM analysis, and FBT can be efficiently implemented directly inside the FPGA in future campaigns, allowing advanced calibration to be performed in real time. We also suggest automatising the full tune-up process of the algorithmic initialisation protocol in preparation for larger-scale operation.

Challenges remain in raising SPAM and control fidelities to far above $\SI{99}{\percent}$ to achieve truly fault tolerant operation. We find that the control process potentially injects errors into the spin readout, which should be addressed to increase the readout fidelity. In the future, incoherent errors can be ameliorated by improving the quality of the Si/SiO$_2$ interface and the SiO$_2$ layer, and reducing the noise level in the experimental setup. We expect that fabrication of SiMOS devices in industrial foundries~\cite{gonzalez-zalba2021scaling,zwerver2022qubits} will bring a reduction in defects and charge impurities~\cite{saraiva2022materials,elsayed2022low}, which will increase qubit coherence times and decrease the required feedback. Faster readout is also desired to reduce the initialisation time and consequently the overall SPAM duration.

Ultimately, the scalability of spin qubits will rely on scalable control techniques, such as the multi-qubit SMART protocol~\cite{seedhouse2021quantum,hansen2021pulse,hansen2022implementation}, in which the qubits are continuously driven by a modulated microwave field. In such schemes, the driving pulses decouple the qubits from noise, and eliminate free precession, during which they are most sensitive to decoherence in the system. Advanced shaping of control pulses can also account for coherent errors arising from miscalibration and parameter drifts.

The engineering challenges in building a fault-tolerant, million-qubit quantum processor remain formidable. One of the most promising pathways to solve them will be the adoption of the extraordinarily successful CMOS chip manufacturing methods. The results presented here show that high-fidelity quantum operations can be achieved in a CMOS-compatible silicon processor, at high enough temperature to realistically permit the operation and integration of classical control circuits, making a truly scalable semiconductor quantum processor a plausible reality in the future.



\section*{Methods}
\setcounter{subsection}{0}

\subsection{Measurement setup}\label{methods:measurement_setup}
The full experimental setup is shown in Extended Data Fig.~\ref{fig:experimental_setup}. The device is measured in a Bluefors XLD400 dilution refrigerator. The device is mounted on the cold finger. Within $T=\SI{1}{\kelvin}$, elevation from the base temperature is achieved by switching on and tuning the heater near the sample. Temperatures above $\SI{1}{\kelvin}$ are attained by reducing the amount of He mixture in the circulation and consequently the cooling power. Temperature control becomes non-trivial above $\SI{1.2}{\kelvin}$ and nonviable above $\SI{1.5}{\kelvin}$.

An external DC magnetic field is supplied by an American Magnetics AMI430 magnet. The magnetic field points in the [110] direction of the Si lattice. DC voltages are supplied with Basel Precision Instruments SP927 LNHR DACs, through DC lines with a bandwidth from 0 to $\SI{20}{\hertz}$. Dynamic voltage pulses are generated with a Quantum Machines OPX and combined with DC voltages via custom voltage combiners at the $\SI{50}{\kelvin}$ stage in the refrigerator. The OPX has a sampling time of $\SI{4}{\nano\second}$. The dynamic pulse lines in the fridge have a bandwidth of 0 to $\SI{50}{\mega\hertz}$, which translates into a minimum rise time of $\SI{20}{\nano\second}$. Microwave pulses are synthesised using a Keysight PSG8267D Vector Signal Generator, with the baseband I/Q and pulse modulation signals from the OPX. The modulated signal spans from $\SI{250}{\kilo\hertz}$ to $\SI{44}{\giga\hertz}$, but is band-limited by the fridge line and the DC block.

The charge sensor comprises a single-island SET connected to a tank circuit for reflectometry measurement. The return signal is amplified by a Cosmic Microwave Technology CITFL1 LNA at the $\SI{4}{\kelvin}$ stage, and a Mini-circuits ZX60-P33ULN+ LNA followed by two Mini-circuits ZFL-1000LN+ LNAs at room temperature. The Quantum Machines OPX generates the tones for the RFSET, and digitises and demodulates the signals after the amplification.

\subsection{Device tune-up}\label{methods:device_tuneup}
We first load the electrons according to the mapping of the double-dot charge configurations over a large range, via lock-in charge sensing measurement~\cite{yang2012orbital} with the RFSET. The measurement can be performed in the physical gate basis by sweeping $V_\mathrm{P1}$ and $V_\mathrm{P2}$, as shown in Extended Data Fig.~\ref{fig:device_tune_up}~a, or in the virtual gate basis by sweeping $V_\mathrm{P1}-V_\mathrm{P2}$ and $V_\mathrm{J}$, as shown in Fig.~\ref{fig:main_fig_1}~a. In the virtual gate basis, voltages of $-0.32V_\mathrm{J}$ and $-0.25V_\mathrm{J}$ are applied on P1 and P2 to compensate the effect of pulsing J. During operation, each dot is loaded with an odd number of electrons, from which the unpaired electron carries the spin information. This is denoted as the $(m+1, n+1)$ charge state in the charge maps, where $m$ and $n$ are even numbers.

The tune-up procedure proceeds with locating the PSB region around the inter-dot charge transition, as indicated by the dashed square in Extended Data Fig.~\ref{fig:device_tune_up}~c and d. The initial PSB search involves loading a mixed spin state in $(m+1, n+1)$, which has some probability of being even-parity ($\ket{\downarrow\downarrow}$ or $\ket{\uparrow\uparrow}$), and subsequently pulsing to a location near the inter-dot charge transition point. Single-shot charge readout is performed before and after reaching the location, and the final readout signal is provided by subtracting the two signals. Except at ultra-low $B_0$, the readout mechanism is dominated by parity readout due to the relatively large Zeeman energy difference between the two qubits~\cite{seedhouse2021pauli}. An even-parity spin state will show up as blockaded in the PSB region, which translates to a lower RF signal compared to that from an unblockaded state. The averaged RF signal therefore indicates the probability of having an even-parity state across multiple shots.

The two-level behaviour in the PSB region is used to perform single-shot spin readout. The readout signal in each shot of experiment is compared to a preset threshold that lies between the two levels, as we see in the readout histograms in Fig.~\ref{fig:main_fig_2}~b. We assign value 1 to a blockaded readout, and value 0 to an unblockaded readout. Finally, we average over all shots to obtain $P_\mathrm{blockade}$ for the statistics.

The speed and exchange level at which we initialise $(m+1, n+1)$ influence the probability of different states by changing the adiabaticity in the transition. At a given bias configuration, a slow adiabatic ramp tends to incur a lowest-energy $\ket{\downarrow\downarrow}$ state (process i in Extended Data Fig.~\ref{fig:device_tune_up}~b and Extended Data Fig.~\ref{fig:device_tune_up}~c), while a fast diabatic ramp can result in the odd-parity states or even $\ket{\uparrow\uparrow}$ (process ii in Extended Data Fig.~\ref{fig:device_tune_up}~b and Extended Data Fig.~\ref{fig:device_tune_up}~d). At millikelvin temperatures, it is possible for this bias in the spin proportions to be large enough for high-fidelity initialisation. The bias is reduced with the increased thermalisation above $\SI{1}{\kelvin}$, but nonetheless visible when comparing the resulting PSB from two vastly different initialisation ramp rates (Extended Data Fig.~\ref{fig:device_tune_up}~e).

In Extended Data Fig.~\ref{fig:device_tune_up}~c to e, an additional latch readout region can also be seen to the bottom-left of the PSB region. While it also provides parity readout, the latch readout involves tunneling to the reservoir and may not be scalable, thus it is not used in the algorithmic initialisation.

Extended Data Fig.~\ref{fig:device_tune_up}~f shows the ESR spectrum as a function of $V_\mathrm{J}$, in which we identify two regimes. At $V_\mathrm{J} < \SI{1.175}{\volt}$, only two transitions pertaining to the driven rotation of the individual qubits are detected. Driven over time, these transitions correspond to the Rabi oscillations in Fig.~\ref{fig:main_fig_1}~e. At $V_\mathrm{J} > \SI{1.175}{\volt}$, where the exchange energy is significant, we see four transitions among the four two-qubit states, corresponding to the controlled rotation (CROT) operations~\cite{huang2019fidelity,noiri2022fast}. 
The layout of the transitions, together with the background signal, reveals the composition of the initialised qubit state, as discussed in the main text. The traces in Fig.~\ref{fig:main_fig_2}~a are taken from such measurements at high $V_\mathrm{J}$. A more scalable two-qubit operation is the electrically pulsed controlled phase operation (CZ)~\cite{watson2018programmable,xue2022quantum}. This is adopted in this work to construct the CZ gate (Extended Data Fig.~\ref{fig:device_tune_up}~g), or the DCZ gate in the main text.

\subsection{Algorithmic initialisation}\label{methods:algorithmic_initialisation}
The algorithmic initialisation protocol, as depicted in Extended Data Fig.~\ref{fig:algorithmic_init}, proceeds as follows:
\begin{enumerate}
    \item[1,] Enter $(m+1, n+1)$ to create two unpaired spins in the double-dot system.
    \item[2,] This results in one of the $\ket{\downarrow\downarrow}$, $\ket{\downarrow\uparrow}$, $\ket{\uparrow\downarrow}$ and $\ket{\uparrow\uparrow}$ states. The probability of creating the ground state $\ket{\downarrow\downarrow}$ decreases as the temperature increases, as the thermal energy becomes comparable or greater than the qubit exchange coupling and the Zeeman energies.
    \item[3,] Ramp to the PSB region for parity readout, and apply a filter that rejects odd-parity states:
    \begin{enumerate}
        \item[i,] If the state is unblockaded and thus determined as odd-parity ($\ket{\downarrow\uparrow}$, $\ket{\uparrow\downarrow}$), the initialisation is restarted.
        \item[ii,] If the state is blockaded and thus determined as even-parity ($\ket{\downarrow\downarrow}$, $\ket{\uparrow\uparrow}$), the initialisation proceeds to the next stage.
    \end{enumerate}
    \item[4,] This results in either $\ket{\downarrow\downarrow}$ or $\ket{\uparrow\uparrow}$, with an increased probability of $\ket{\downarrow\downarrow}$ from Step 3. We calibrate the CZ gate at this stage, either from the exchange-induced splitting of the ESR transitions (Extended Data Fig.~\ref{fig:device_tune_up}~f), or the CZ oscillations (Extended Data Fig.~\ref{fig:device_tune_up}~g).
    \item[5,] A zCNOT gate is performed to convert $\ket{\uparrow\uparrow}$ into $\ket{\uparrow\downarrow}$, leaving $\ket{\downarrow\downarrow}$ unchanged. The construction of the zCNOT gate in this work is shown in Extended Data Fig.~\ref{fig:device_tune_up}~g.
    \item[6,] Ramp to the PSB region for parity readout, and apply a filter that rejects odd-parity states:
    \begin{enumerate}
        \item[i,] If the state is unblockaded and thus determined as $\ket{\uparrow\downarrow}$, the initialisation is restarted.
        \item[ii,] If the state is blockaded and thus determined as $\ket{\downarrow\downarrow}$, the initialisation is determined to be completed.
    \end{enumerate}
    \item[7,] The resulting state is purely $\ket{\downarrow\downarrow}$.
\end{enumerate}

The if conditions above are implemented using real-time logics in the FPGA.

The protocol can be adapted to prepare any other state in the parity basis. $\ket{\uparrow\downarrow}$ and $\ket{\downarrow\uparrow}$ can be prepared from $\ket{\downarrow\downarrow}$ with a microwave $\pi$ pulse on Q1 and Q2. $\ket{\uparrow\uparrow}$ can be prepared by replacing the zCNOT with CNOT in the algorithm.

We apply the algorithmic initialisation in a wide range of $B_0$ from $\SI{1}{\tesla}$ down to $\SI{25}{\milli\tesla}$. At low $B_0$ such as $\SI{85}{\milli\tesla}$, the resulting state is highly mixed without the algorithmic initialisation, as we see in Extended Data Fig.~\ref{fig:algorithmic_init}~b. Step 1 to 3 of the algorithmic initialisation is still highly effective in creating an even-parity state, and the full algorithmic initialisation is successful in removing $\ket{\downarrow\downarrow}$. The small transition amplitude in the ESR spectra is now dominated by not only exchange, but also the non-standard qubit control and readout at low $B_0$ (see Extended Data Fig.~\ref{fig:B0_dependence}~f).

It is also important to assess the time cost for the algorithmic initialisation, since it involves multiple control and readout iterations. The table in Extended Data Fig.~\ref{fig:algorithmic_init}~b breaks down the time spent on control and readout. We see that the readout integration time $t_\mathrm{integration}$ dominates the time consumption. At $B_0 = \SI{85}{\milli\tesla}$ and $T=\SI{1}{\kelvin}$, the full algorithmic initialisation takes a average of around 3 iterations, which totals around $\SI{150}{\micro\second}$. Evaluating this in the context of different $B_0$ and temperatures, we obtain the dependence shown in Extended Data Fig.~\ref{fig:algorithmic_init}~c and d. At ultra-low $B_0$ where a reduction in the control and readout fidelity is seen, $N_\mathrm{iteration}$ decreases, possibly because the system deviates from the parity basis. Higher $B_0$ provides a larger qubit energy, increasing the likelihood of obtaining a $\ket{\downarrow\downarrow}$ state after the load ramp and reducing $N_\mathrm{iteration}$. In a similar manner, $N_\mathrm{iteration}$ also increases with higher temperatures. At $B_0$ above $\SI{1}{\tesla}$, the onset of excited state level crossings enhances spin randomisation after the load ramp, and thus more $N_\mathrm{iteration}$ is required. We expect that $N_\mathrm{iteration}$ may be reduced by incorporating corrective control based on measurement~\cite{philips2022universal,kobayashi2023feedback} to accelerate the polarisation towards the target state.

\subsection{SPAM error analysis with repeated readout}\label{methods:SPAM_analysis}

A more comprehensive SPAM error analysis utilises machine learning of the increased statistics from multiple measurements. The experimental sequence consists of initialisation followed by repeated parity readout, which results in a series of binary measurement outcomes $m_1, m_2, ..., m_{n}$, where $m_i \in \{\text{even parity} = 0, \text{odd parity} = 1 \}$. This initialisation-$(\mathrm{measurement})^n$ sequence is performed 1000 shots. 

A hidden Markov model (HMM) can describe this series of measurements formalism where the true, but hidden, spin state $s_1, s_2, ..., s_{n}$ follows the Markov chain and the measurement outcomes,  $m_i$, are probabilistically related to the underlying spin state. Three different tensors completely determine HMMs:
\begin{enumerate}
    \item[1,] A start probability vector, $\vec{\Pi}$, encoding the initialising probabilities in each spin state. 
    \item[2,] A transition probability matrix, $\mathbf{A}$, encoding the probabilities of transiting between spin states during measurements. 
    \item[3,] A measurement probability matrix, $\mathbf{\Theta}$, encoding the probability of the measurement outcomes conditioned on the current hidden spin state. 
\end{enumerate}

To find the likely HMM model for a given set of data, we perform expectation maximisation whereby we maximise the marginal likelihood, which is dependent on the marginalised hidden spin state, such that:
\begin{align}\label{eq:marginal_likelihood}
	L(\vec{\Pi}, \mathbf{A}, \mathbf{\Theta}; \vec{m}) &:= p(\vec{m}| \vec{\Pi}, \mathbf{A}, \mathbf{\Theta}) \nonumber 
	\\ &= \int p(\vec{s}, \vec{m}| \vec{\Pi}, \mathbf{A}, \mathbf{\Theta}) d\vec{s}.
\end{align}
For HMM models there exists the Baum-Welch algorithm which can perform this expectation maximisation via an iterative update rule, without the need for back propagation of gradients~\cite{rabiner1986}. 

We use the Cramer-Rao bound to quantify the level of uncertainty in these parameters when fitted by expectation maximisation~\cite{cramer1999}. The Cramer-Rao bound states that if $\text{est}_{\vec{\theta}}(\vec{m})$ is an unbiased estimate of the parameters $\vec{\theta} := (\vec{\Pi}, \mathbf{A}, \mathbf{\Theta})$ given the data $\vec{m}$, such as that produced by expectation maximisation, then: 
\begin{equation}\label{eq:CR_bound}
	\text{cov}_{\vec{\theta}} \left(\text{est}_{\vec{\theta}}(\vec{m}) \right) \geq I \left(\vec{\theta}; \vec{m} \right)^{-1},
\end{equation}
where $I(\vec{\theta}; \vec{y})_{ij} = -\partial^2 \log L(\vec{\theta}; \vec{m}) / \partial \theta_i \partial \theta_j$, the Fisher information matrix. Therefore, we can obtain lower bounds on each parameters uncertainty from the diagonal elements of the inverse of the Fisher information matrix. We used the Forward-Backward algorithm to compute the marginal likelihood defined in equation~(\ref{eq:marginal_likelihood}) needed to compute the Fisher information matrix.

Finally, we use the Viterbi algorithm to compute the most likely set of true spin states which gave rise to the set of measurements given a set of model parameters~\cite{rabiner1986, viterbi1967error}.

\subsection{Crosstalk correction}\label{methods:crosstalk}

The relatively small $\Delta E_\mathrm{Z}$ even at higher $B_0$ requires cancellation of crosstalk between the two qubits, that is, the effect on the the other qubit when one qubit is being driven. This can be addressed to the first order by considering the following aspects.

To cancel off-resonance driving, we enforce
\begin{equation}\label{eq:off_res_driving}
\sqrt{\Delta{E_\mathrm{Z}^2}+f_\mathrm{Rabi}^2}=Nf_\mathrm{Rabi},
\end{equation}
where $\Delta{E_\mathrm{Z}}$ is the Zeeman frequency difference between the qubits, $f_\mathrm{Rabi}$ is the Rabi frequency of the target qubit, and $N=4, 8, 12, ...$~. Consequently, each $\pi/2$ microwave pulse on the target qubit incurs a full $2\pi{N}$ off-resonance rotation on the ancilla qubit, as exemplified in Extended Data Fig.~\ref{fig:crosstalk}~a. Failure to cancel the off-resonance driving can result in large errors under parity readout, as shown in Fig.~\ref{fig:crosstalk}~b. With $N=4$, this cancellation criterion dictates the fastest Rabi possible and is therefore expected to limit the single-qubit gate fidelities, especially at low $B_0$ where $\Delta{E_\mathrm{Z}}$ is small. The full set of $f_\mathrm{Rabi}$ used for single-qubit RB at different $B_0$ is shown in Extended Data Fig.~\ref{fig:crosstalk}~c. In this case, we can alternatively execute $\mathrm{X}(\pi/2)$ as a $3\pi/2$ gate for faster driving at the cost of redundancy. We implemented this with the 3- and 5-electron qubit at $\SI{0.1}{\tesla}$, $\SI{1.2}{\kelvin}$ in Fig.~\ref{fig:main_fig_3}~d.

In two-qubit sequence runs, it is also necessary to correct AC Stark shift by an amount of
\begin{equation}\label{eq:AC_Stark_shift}
\frac{f_\mathrm{Rabi}^2}{2\Delta{E_\mathrm{Z}}},
\end{equation}
apart from cancelling the off-resonance driving. Extended Data Fig.~\ref{fig:crosstalk}~d measures the AC Stark shift on an ancilla qubit by preparing it on the equator, driving it off-resonantly and projecting the phase. Before correction, the AC Stark shift is seen as the linear fringes that correspond to the phase accumulation given by equation \ref{eq:AC_Stark_shift}.

We note that the above cancellation of crosstalk does not prevent it from incurring errors. The perturbation on the ancilla qubit induces decoherence. At ultra-low $B_0$ where $\Delta E_\mathrm{Z}$ becomes diminishing, higher-order crosstalk terms cannot be neglected, and the control of individual qubits becomes unmanageable. However, these problems are circumvented in the SMART control scheme, which addresses all the qubits simultaneously. 

\subsection{Randomised benchmarking}\label{methods:RB}

Single-qubit randomised benchmarking (RB) sequences for Fig.~\ref{fig:main_fig_3}~d--e are constructed from elementary
$\pi/2$ gates [$\mathrm{X}(\pi/2)$, $\mathrm{Z}(\pi/2)$, $-\mathrm{X}(\pi/2)$, $-\mathrm{Z}(\pi/2)$], $\pi$ gates [$\mathrm{X}(\pi)$, $\mathrm{Z}(\pi)$] and an $\mathrm{I}$ gate. Each Clifford gate contains one physical elementary gate on average, excluding the virtual $\mathrm{Z}(\pi/2)$ and $\mathrm{Z}(\pi)$ gates.

Two-qubit RB sequences for Fig.~\ref{fig:main_fig_4}~b are constructed from single-qubit elementary
$\pi/2$ gates [$\mathrm{X}_1(\pi/2)$, $\mathrm{Z}_1(\pi/2)$, $\mathrm{X}_2(\pi/2)$, $\mathrm{Z}_2(\pi/2)$] for Q1 and Q2, and a two-qubit elementary gate $\mathrm{DCZ}$. Each Clifford gate contains 1.8 single-qubit elementary gates and 1.5 two-qubit elementary gates on average. All gates are sequentially executed, which means Q1 idles while $\mathrm{X}_2(\pi/2)$ or $\mathrm{Z}_2(\pi/2)$ takes place, and the same for Q2. The generated random sequences are used in both RB and FBT. In the case of IRB, we incorporate an interleaved $\mathrm{DCZ}$ gate between adjacent Clifford gates. The experimental implementation and the analysis protocol are shown in Extended Data Fig.~\ref{fig:RB_and_FBT_raw}~a--b, and the IRB results are shown in Extended Data Fig.~\ref{fig:RB_and_FBT_raw}~c.

We then fit the RB decay curve to the formula~\cite{yang2019silicon,huang2019fidelity}
\begin{equation}\label{eq:RB_fit}
    ae^{-{(bx)}^c}+d,
\end{equation}
from which $1-0.5b$ gives the Clifford fidelity in single-qubit RB, and $1-0.75b$ gives the Clifford fidelity in two-qubit RB. $c$ represents the decay exponent and reflects the error Markovianity. $a$ is subjected to the readout fidelity, and $d$ is close to 0.5.

It should be noted that spin relaxation, excitation, incorrect rotation or slow drifts in charge readout can obscure long sequence runs. Under our operating conditions at $T=\SI{1}{\kelvin}$, the longest RB sequences in our experiment reach an average of 1400  elementary gates, taking up to several hundreds of microseconds. This is well within the $T_1$ we measure. We measure the decay in the $\mathrm{+ZZ}$ (no operation before parity readout) projection and the $\mathrm{-ZZ}$ projection ($\pi$ pulse on a single qubit before parity readout) after the recovery gate (see Supplementary Fig.~\ref{fig:long_RB_dual_projection}).

\subsection{Fast Bayesian tomography}\label{methods:FBT}
Fast Bayesian tomography~\cite{evans2022fast,su2023characterizing} (FBT) is an agile gate set process tomography protocol that can self-consistently reconstruct all gate set process matrices based on prior calibration. In principle, FBT learns and updates the model using the gate sequence information and its experimental outcome. In this work, we feed FBT with the variable-length two-qubit RB sequences and the corresponding experimental data. Clifford gates in the RB sequences are decomposed into their elementary gate implementation of $\mathrm{X}_1(\pi/2)$, $\mathrm{Z}_1(\pi/2)$, $\mathrm{X}_2(\pi/2)$, $\mathrm{Z}_2(\pi/2)$, and $\mathrm{DCZ}$. The RB experiments at $T=\SI{0.1}{K}$ and $T=\SI{1}{K}$ run through 32000 and 26000 sequences respectively, sufficient for FBT to reliably reconstruct the error channels. We feed the native parity readout results directly to FBT, without converting them to the standard two-qubit measurement basis.

To initiate the FBT analysis, we must bootstrap the model from educated guesses to help the analysis converge with a finite amount of experiments. Here, we do this by injecting guessed fidelity numbers as introduced in~\cite{evans2022fast,su2023characterizing}. FBT models each noisy gate $\tilde{G}$ as the product of the noise channel $\tilde{G}=\Lambda G$ and the ideal gate $G$, where the noise channel is linearised about $I$ by expressing it as $\Lambda=I+\varepsilon$. Each update of the FBT analysis is essentially on the statistics of the noise channel residuals, $\varepsilon$. Extended Data Fig.~\ref{fig:RB_and_FBT_raw}~d shows the reconstructed Pauli transfer matrices (PTMs) of the $\mathrm{DCZ}$ gate. Supplementary Fig.~\ref{fig:error_generators} shows the reconstructed noise channel residuals of the three physical elementary gates $\mathrm{DCZ}$, $\mathrm{X}_1(\pi/2)$, and $\mathrm{X}_2(\pi/2)$ at $T=\SI{0.1}{\kelvin}$ and $T=\SI{1}\kelvin$.

Since FBT does not guarantee that the reconstructed channels are physical or flag any gauge ambiguity, we perform CPTP projection and gauge optimisation over the whole gate set at the output stage.

\subsection{Error taxonomy with pyGSTi}\label{methods:error_taxonomy}
Error taxonomy for FBT can be achieved by converting the noise channels ($\Lambda$) for each gate to their error generator($\mathds{L}$) using the following relationship:
\begin{align}\label{eq:error_generator}
    G = \Lambda G_0 = e^{\mathds{L}} G_0,
\end{align}
where $G$ is the estimated noisy gate, and $G_0$ is the ideal gate.

Using pyGSTi~\cite{nielsen2022pyGSTio, blume-kohout2022a}, we project $\mathds{L}$ into the subspace of Hamiltonian and Stochastic errors, extracting the coefficients of each elementary error generator. We perform this analysis on each of the gates [$\mathrm{DCZ}$, $\mathrm{X}_1(\pi/2)$, and $\mathrm{X}_2(\pi/2)$], for both temperatures of $\SI{0.1}{\kelvin}$ and $\SI{1}\kelvin{}$. The coefficients of the elementary error generators are represented in the Pauli basis and plotted in Supplementary Fig.~\ref{fig:error_channels}. The five largest components of the Hamiltonian and Stochastic errors for the $\mathrm{DCZ}$ gate are summarised in Fig.~\ref{fig:main_fig_4}~c.

We also estimate the generator or entanglement infidelity $1 - \mathcal{F}_\mathrm{ent}$ based on these error coefficients, given by~\cite{blume-kohout2022a}:
\begin{align}\label{eq:error_coefficients}
    1 - \mathcal{F}_\mathrm{ent} \approx \sum_P s_P + \sum_P h_P^2,
\end{align}
where the sum is performed over the extracted coefficients and $P$ denotes non-identity Pauli elements. The approximation is validated by the domination of Hamiltonian errors over stochastic errors in magnitude. To obtain the average gate fidelities ($\mathcal{F}_\mathrm{avg}$), which are the quantities quoted based on IRB and FBT measurements, it can be connected to $\mathcal{F}_\mathrm{ent}$ in the following way~\cite{horodecki1999general}:
\begin{align}\label{eq:entanglement_fidelity}
    \mathcal{F}_\mathrm{avg} = \frac{d\cdot\mathcal{F}_\mathrm{ent} + 1}{d+1},
\end{align}
where $d$ is the dimension of the Hilbert space (4 for a two-qubit system). This means that generally stochastic errors contribute more to the gate infidelities, even in the case where the magnitudes of the Hamiltonian errors are larger.

\section*{Acknowledgements}
We acknowledge technical support from Alexandra Dickie. We acknowledge technical GST discussions with Corey Ostrove and Robin Blume-Kohout. We acknowledge support from the Australian Research Council (FL190100167 and CE170100012), the U.S. Army Research Office (W911NF-23-10092), the U.S. Air Force Office of Scientific Research (FA2386-22-1-4070), and the NSW Node of the Australian National Fabrication Facility. The views and conclusions contained in this document are those of the authors and should not be interpreted as representing the official policies, either expressed or implied, of the Army Research Office, the U.S. Air Force or the U.S. Government. The U.S. Government is authorised to reproduce and distribute reprints for Government purposes notwithstanding any copyright notation herein. J.Y.H., R.Y.S., M.K.F., S.S., J.D.C., I.H., and A.E.S. acknowledge support from Sydney Quantum Academy.

\section*{Author contributions}
J.Y.H., R.Y.S., A.S., A.L., A.S.D., and C.H.Y. designed the experiments. J.Y.H. performed the experiments under A.S., A.L., A.S.D., and C.H.Y.'s supervision. W.H.L. and F.E.H. fabricated the device under A.S.D.'s supervision on enriched $^{28}$Si wafers supplied by N.V.A., H.-J.P., and M.L.W.T.. S.S. designed the RFSET setup. W.G., N.D.S., S.S., E.V., and A.L. contributed to the experimental hardware setup. W.G., N.D.S., and S.S. contributed to the experimental software setup. T.T. assisted with the two-qubit randomised sequence generation. B.V.S. and B.S. performed the SPAM error analysis with machine learning under A.S. and N.A.'s supervision. R.Y.S. performed the noise spectroscopy analysis. I.H., A.E.S., and C.H.Y. assisted with the SMART protocol implementation. R.Y.S. performed the FBT analysis under T.T., A.S., and S.D.B.'s supervision. M.K.F. performed the subsequent error generator analysis with pyGSTi under A.S.'s supervision. R.Y.S., W.H.L., M.K.F., W.G., N.D.S., T.T., J.D.C., C.C.E., S.D.B., A.M., A.S., A.L., A.S.D., and C.H.Y. contributed to the discussion, interpretation and presentation of the results. J.Y.H., R.Y.S., M.K.F., B.V.S., F.E.H., S.D.B., A.L., A.S.D., and C.H.Y. wrote the manuscript, with input from all co-authors.

\section*{Corresponding authors}
Correspondence to J.Y.H., A.S.D., or C.H.Y..\\\\

\section*{Competing interests}
A.S.D. is the CEO and a director of Diraq Pty Ltd. W.H.L., W.G., N.D.S., T.T., E.V., C.C.E., F.E.H., A.S., A.L., A.S.D., and C.H.Y. declare equity interest in Diraq Pty Ltd. J.Y.H., A.S., and C.H.Y. are inventors on a patent related to this work (AU provisional application 2023902138) filed by Diraq Pty Ltd with a priority date of $3^\mathrm{rd}$ July 2023.

\section*{Data availability}
All data of this study will be made available in an online repository.

\section*{Code availability}
The analysis codes that support the findings of the study are available from the corresponding authors on reasonable request.

\setcounter{figure}{0}
\setcounter{table}{0}
\captionsetup[figure]{name={\bf{Extended Data Fig.}},labelsep=line,justification=centerlast,font=small}
\captionsetup[table]{name={\bf{Extended Data Table}},labelsep=period,justification=centerlast,font=small}

\onecolumngrid
\vfill

\begin{table*}[h]
\centering
\caption{\textbf{Key metrics of the two-qubit processor.}}
\label{table:fidelities}
\begin{tabularx}{0.93\linewidth}{|c|c|>{\centering}p{6.25em}|>{\centering}p{6.25em}|>{\centering}p{6.25em}|>{\centering}p{7.25em}|>{\centering}p{5.45em}|}
\cline{1-7}
\multicolumn{2}{|c|}{Operating condition} & \multicolumn{5}{c|}{Fidelity~($\unit{\percent}$)}
\cr \cline{1-7}
External magnetic field & Temperature & Initialise even & Readout even & Readout odd & 1Q Clifford gate & DCZ gate
\cr \cline{1-7}
\multirow{2}{*}{\shortstack{$B_0=\SI{0.79}{\tesla}$\\($f_\mathrm{Rabi}=\SI{1.84}{\mega\hertz}$)}} & $T=\SI{0.1}{\kelvin}$ & $99.40\pm 0.25$ & $99.69\pm 0.07$ & $96.79\pm 0.12$ & - & $99.15\pm 0.13$
\cr \cline{2-7}
& $T=\SI{1}{\kelvin}$ & $99.34\pm 0.27$ & $99.34\pm 0.08$ & $96.15\pm 0.44$ & $99.60\pm 0.01$ & $98.92\pm 0.67$
\cr \cline{1-7}
\multirow{2}{*}{\shortstack{$B_0=\SI{0.4}{\tesla}$\\($f_\mathrm{Rabi}=\SI{2.6}{\mega\hertz}$)}} & $T=\SI{0.14}{\kelvin}$ & - & - & - & $99.89\pm 0.01$ & -
\cr \cline{2-7}
& $T=\SI{1}{\kelvin}$ & - & - & - & $99.85\pm 0.01$ & -
\cr \cline{1-7}
\end{tabularx}
\\[1em]
\begin{tabularx}{0.93\linewidth}{|c|c|c|c|c|c|c|}
\cline{1-7} \multicolumn{2}{|c|}{Operating condition} & Relaxation time~($\unit{\milli\second}$) & \multicolumn{2}{c|}{Dephasing time~($\unit{\micro\second}$)} & \multicolumn{2}{c|}{Error bias}
\cr \cline{1-7} External magnetic field & Temperature & $T_1$ & $T_2^{*}$ & $T_2^\mathrm{Hahn}$ & $T_1/T_2^*$ & $T_2^\mathrm{Hahn}/T_2^*$
\cr \cline{1-7} \multirow{2}{*}{$B_0=\SI{0.79}{\tesla}$} & $T=\SI{0.14}{\kelvin}$ & $331.29\pm 78.00$ & $3.44\pm 0.13$ & $76.86\pm 17.08$ & $961305\pm 262661$ & $22\pm 6$
\cr \cline{2-7} & $T=\SI{1}{\kelvin}$ & $9.29\pm 3.99$ & $2.32\pm 0.19$ & $33.26\pm 3.38$ & $4004\pm 2048$ & $14\pm 3$
\cr \cline{1-7} \multirow{2}{*}{$B_0=\SI{0.4}{\tesla}$} & $T=\SI{0.14}{\kelvin}$ & $19.48\pm 3.64$ & $3.60\pm 0.14$ & $95.85\pm 2.41$ & $5411\pm 1207$ & $27\pm 2$
\cr \cline{2-7} & $T=\SI{1}{\kelvin}$ & $7.74\pm 1.20$ & $2.32\pm 0.10$ & $32.65\pm 1.32$ & $3336\pm 661$ & $14\pm 1$
\cr \cline{1-7}
\end{tabularx}
\end{table*}

\newpage

\begin{figure*}[ht!]
    \includegraphics[angle = 0]{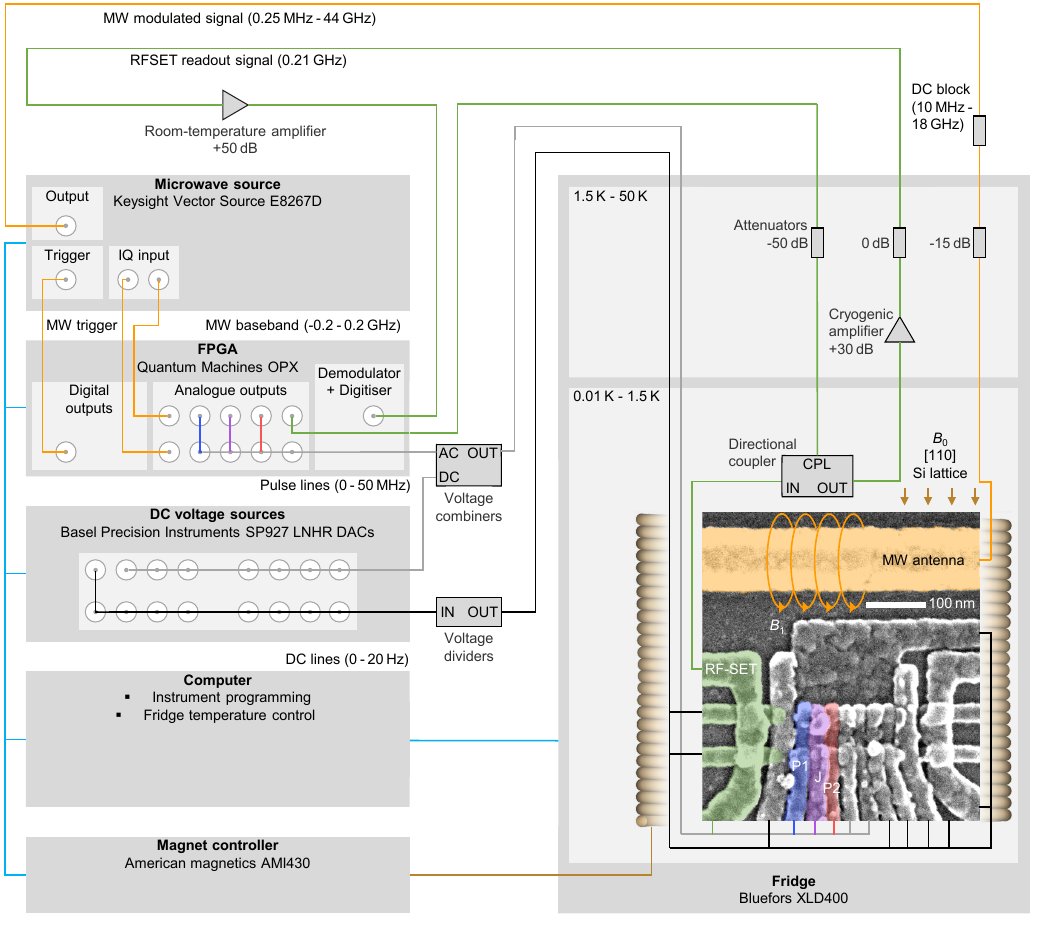}
    \caption{\textbf{Full experimental setup.} 
    }
    \label{fig:experimental_setup}
\end{figure*}

\begin{figure*}[ht!]
    \includegraphics[angle = 0]{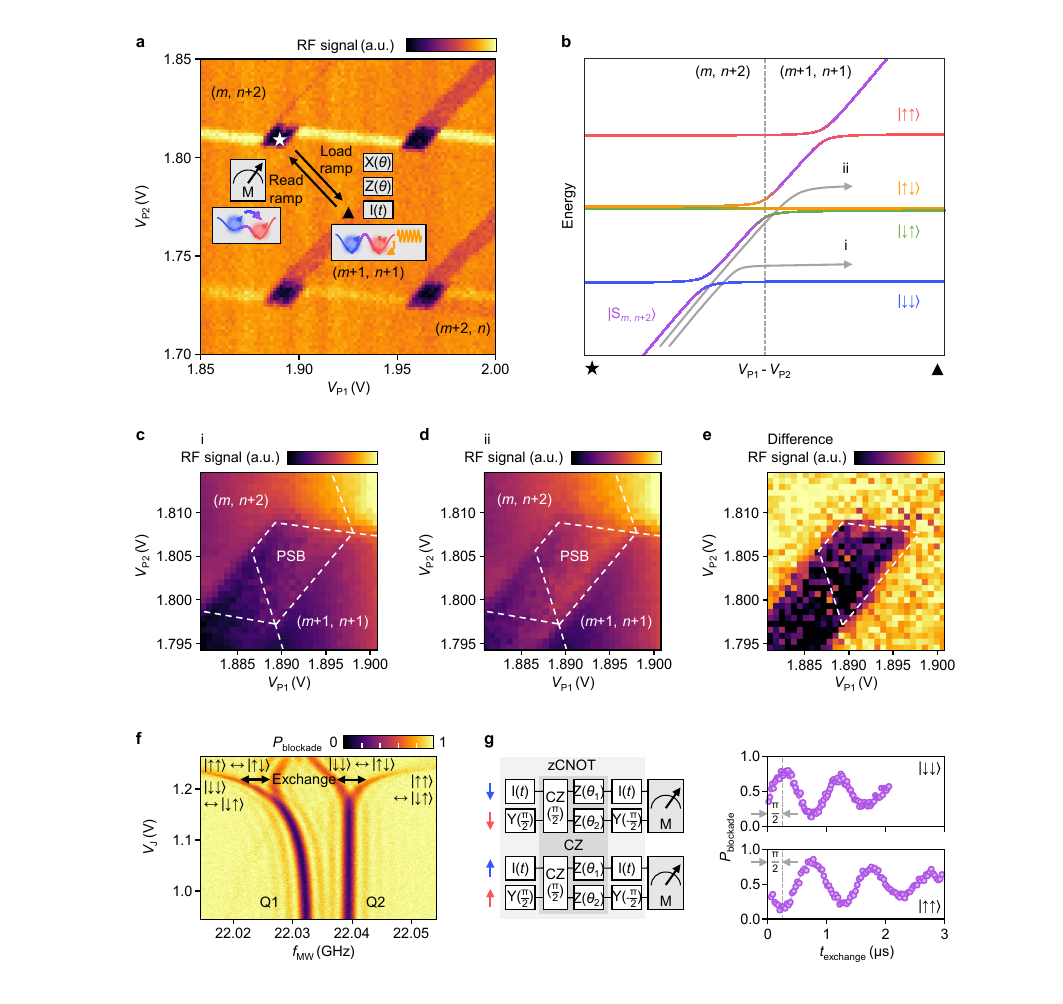}
    \caption{\textbf{Device tune-up.} 
    \textbf{a}, Charge stability diagram as a function of $V_\mathrm{P1}$ and $V_\mathrm{P2}$, showing the operation regime. The readout and control point are labelled with star (\scalebox{1.1}{$\star$}) and triangle (\scalebox{0.75}{$\blacktriangle$}).
    \textbf{b}, Schematic energy diagram of a double-dot system across the inter-dot transition between the $(m, n+2)$ and $(m+1, n+1)$ charge state, where $m$ and $n$ are even numbers. $(m+1, n+1)$ represents a charge state with an unpaired electron spin in each of the dots. The two arrows labelled i and ii refer to two possible loading mechanisms, i being the more adiabatic process.
    \textbf{c}, Averaged signal from 50 shots of charge readout around the $(m+1, n+1)$-$(m, n+2)$ transition, showing partial blockade. The electrons are initialised into $(m+1, n+1)$ via an adiabatic ramp.
    \textbf{d}, Averaged signal from 50 shots of charge readout around the $(m+1, n+1)$-$(m, n+2)$ transition, with the $(m+1, n+1)$ state diabatically initialised. The blockade is visibly weaker.
    \textbf{e}, Different between the readout signals in c and d.
    \textbf{f}, ESR spectrum as a function of $V_\mathrm{J}$, showing the exchange opening up at $V_\mathrm{J}$ above $\SI{1.1}{\volt}$.
    \textbf{g}, Construction of a zCNOT gate, and calibration of the encompassed CZ gate. The CZ gate consists of a CZ operation with $\pi/2$ duration, followed by single-qubit virtual phase corrections to account for the Stark shifts.
    }
    \label{fig:device_tune_up}
\end{figure*}

\begin{figure*}[ht]
    \centering
    \includegraphics[angle = 0]{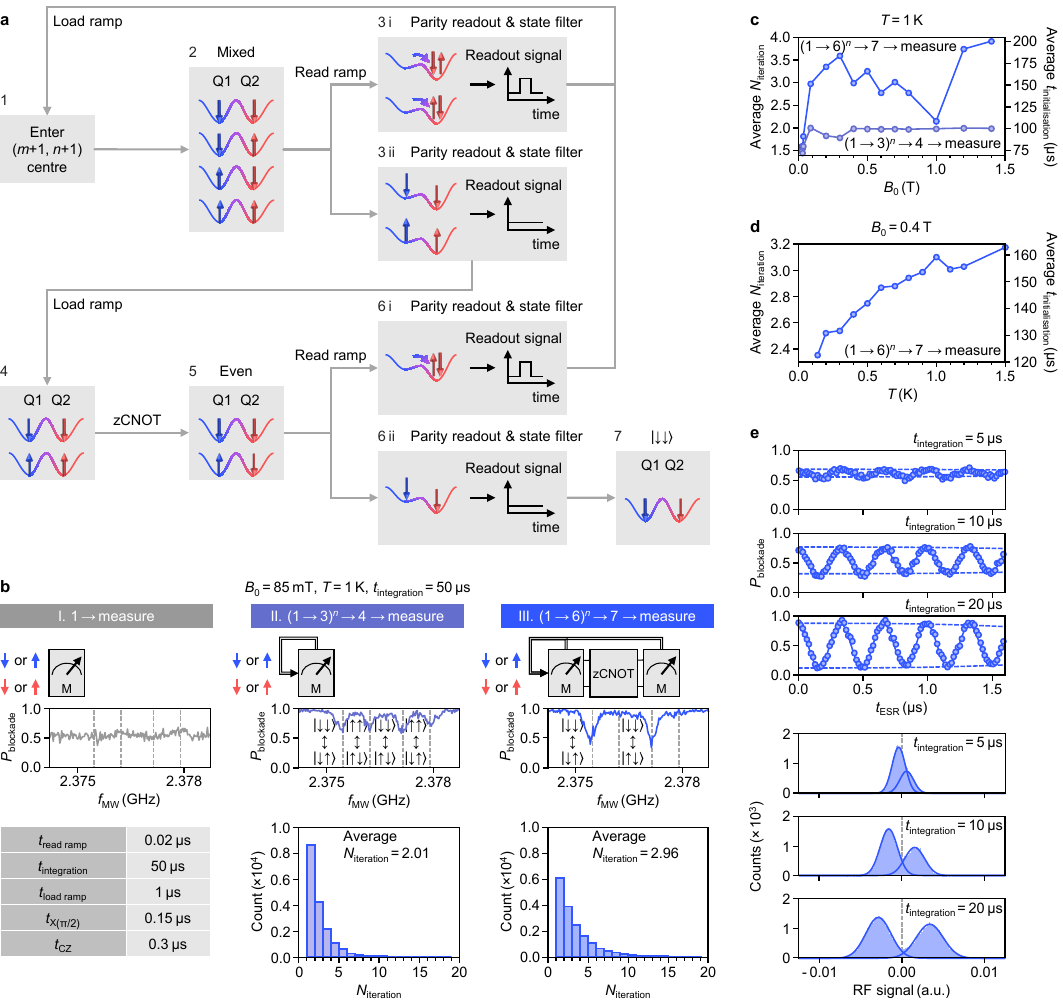}
    \caption{\textbf{Two-qubit algorithmic initialisation.}
    \textbf{a}, Full protocol of the algorithmic initialisation of a two-qubit system, based on parity readout.
    \textbf{b}, Experiments with various extents of the algorithmic initialisation and the corresponding ESR spectra. The data are taken at $T=\SI{1}{\kelvin}$ and $B_0=\SI{85}{\milli\tesla}$, where the thermal energy is 8 times greater than the qubit energies. The table shows the nominal duration for each part of the operation, used in this work. Initialisation with only Step 1 corresponds to the conventional ramped initialisation. The first part of the algorithmic initialisation repeats Step 1 to 3 until an even-parity state is detected. The full algorithmic initialisation repeats Step 1 to 6, in order to detect if the state is solely $\ket{\downarrow\downarrow}$. With the partial or the full algorithmic initialisation, measured with 20000 shots each, we record the statistics on the numbers of iterations, $N_\mathrm{iteration}$, and evaluate the respective average $N_\mathrm{iteration}$.
    \textbf{c}, Average $N_\mathrm{iteration}$ as a function of $B_0$ at $T=\SI{1}{\kelvin}$. Taking the duration amounts from b into account, the average time cost for initialisation, $t_\mathrm{initialisation}$ is estimated.
    \textbf{d}, The quantities in c as a function of temperature at $B_0 = \SI{0.4}{\tesla}$.
    \textbf{e}, Rabi oscillations and charge readout histograms with different amounts of readout integration time, $t_\mathrm{integration}$, at $B_0 = \SI{0.4}{\tesla}$, $T=\SI{1}{\kelvin}$. With short $t_\mathrm{integration}$, the Rabi amplitude becomes limited by the charge readout instead. This may be improved by more advanced readout techniques, such as a double-island SET~\cite{huang2021a} operating at RF or gate dispersive readout~\cite{crippa2019gate}. 
    }
    \label{fig:algorithmic_init}
\end{figure*}

\begin{figure*}
    \centering
    \hspace{-0.6cm}
    \includegraphics[angle = 0]{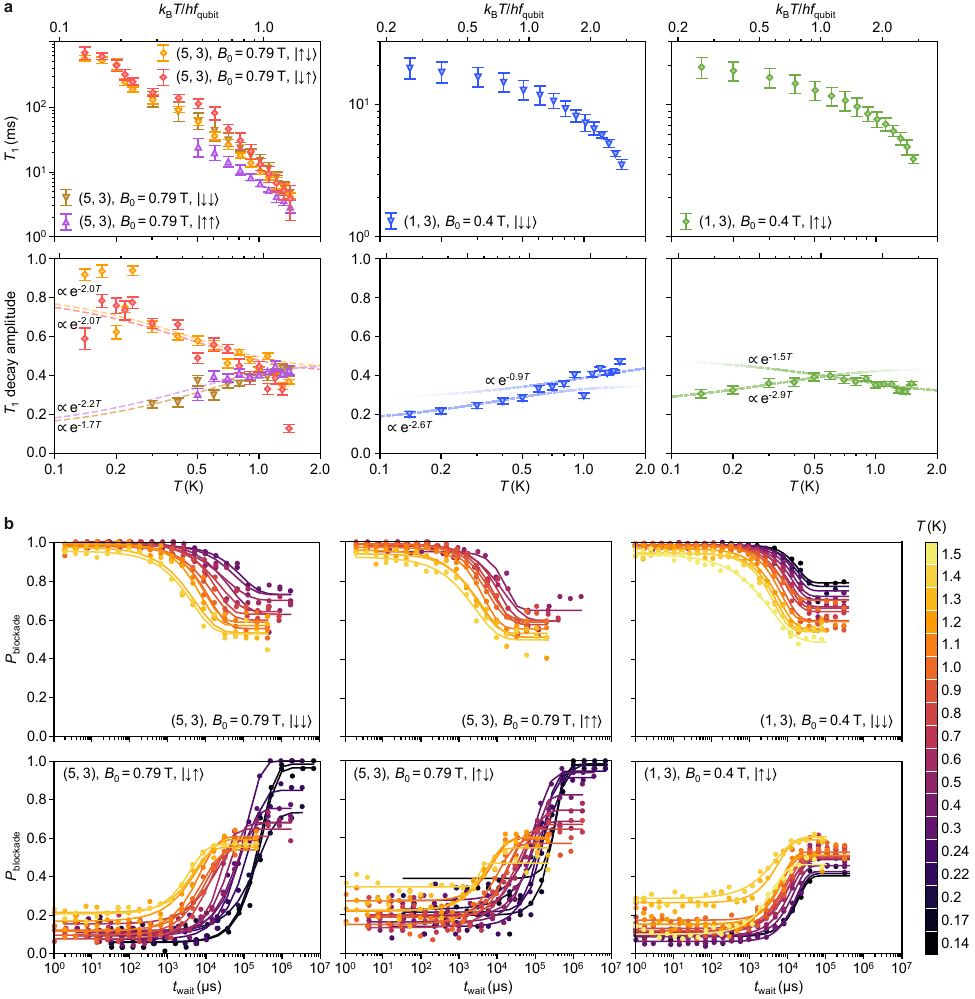}
    \caption{\textbf{$T_1$ processes and temperature dependence.}
    \textbf{a}, The characteristic time of spin relaxation, $T_1$ and the decay amplitude for various two-qubit states, as a function of temperature. We recognise the presence of different relaxation mechanisms at low temperatures, as described in the main text. Here we also look at the evolution of the decay amplitudes, defined as the difference in $P_\mathrm{blockade}$ between the starting point and decay equilibrium. At low temperatures where the relaxation to low-energy states dominates, the decay reaches an equilibrium with mostly $\ket{\downarrow\downarrow}$. With even-parity initialisation, the decay amplitude should be well below 0.5. With odd-parity initialisation, the decay amplitude should be well above 0.5. At high temperatures where the thermal energy becomes comparable or greater than the qubit energy, the decay equilibrium is a mixed state, and $P_\mathrm{blockade}$ tends towards 0.5. Therefore, the decay amplitude reduces as the temperature increases, following an $e^{-k_\mathrm{B}T}$-like reduction, until the degradation of readout starts to dominate. This trend is apparent in the $(5,3)$ state, but becomes more convoluted in $(1,3)$, possibly due to lower-lying excited states. Although $T_1$ is not the limiting time scale in this temperature range, we recognise the rich physical processes behind relaxation revealed in this work additional to the previous results~\cite{petit2018spin,yang2020operation}, and their potential impact on longer or higher-temperature operation in the future.
    \textbf{b}, Measured and fitted relaxation decay curves. Since all the decay curves are one-way, they are fit to a single formula $ae^{-(t/T_1)^c}+d$, where $a$, $c$ and $d$ are the decay amplitude, exponent and equilibrium. Although fluctuations in the readout level is inevitable before RFSET feedback takes place at the end of each shot, the two-level separation in the charge readout is sufficiently large to maintain an overall correct readout level (Supplementary Fig.~\ref{fig:T1_histograms}).
    }
    \label{fig:T1s_raw}
\end{figure*}

\begin{figure*}
    \centering
    \includegraphics[angle = 0]{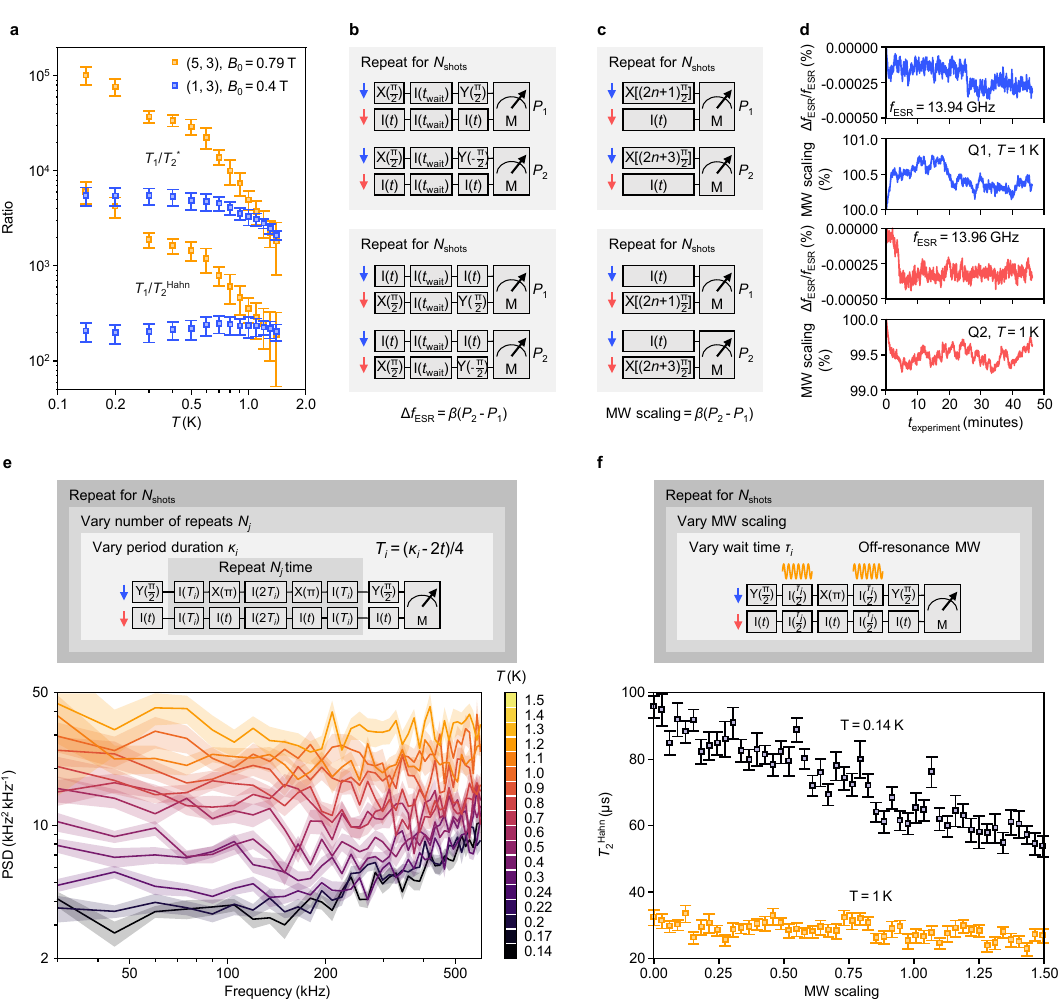}
    \caption{\textbf{Single-qubit temperature dependence, stability, and noise characteristics.}
    \textbf{a}, Ratio of $T_\mathrm{1}$ to $T_\mathrm{2}$ as a function of temperature in different regimes. This ratio indicates the amount of bias in the proportion of depolarisation errors to that of dephasing errors. A large variation in the bias and its temperature dependence is seen at temperatures below $\SI{1}{\kelvin}$, whereas these metrics become similar above $T = \SI{1}{\kelvin}$. At this point, $T_\mathrm{1}/T_\mathrm{2}^\mathrm{*}$ shows a high-order roll-off. However, the temperature dependence is weaker when echoing is incorporated, as seen in $T_\mathrm{1}/T_\mathrm{2}^\mathrm{Hahn}$. The overall $T_\mathrm{1}/T_\mathrm{2}$ biases remain above 100 within $T=\SI{1.5}{\kelvin}$.
    \textbf{b}, Sequences for tracking slow changes in $f_\mathrm{ESR}$ over a long time with respect to $T_2$.~\cite{zhao2019single}
    \textbf{c}, Sequences for tracking the amount of adjustment in microwave power to maintain a constant $f_\mathrm{Rabi}$ over time~\cite{gilbert2022on}. $P_1$, $P_2$ correspond to the different projection outcomes, and $\beta$ is a conversion factor.
    \textbf{d}, Results of a and b at $B_0 = \SI{0.5}{\tesla}$ and $T=\SI{1}{\kelvin}$. $P_1$, $P_2$ correspond to the different projection outcomes, and $\beta$ is a conversion factor.
    \textbf{e}, Sequence for the noise spectroscopy based on the Carr-Purcell-Meiboom-Gill (CPMG) protocol~\cite{cywinski2008enhance,alvarez2011measuring,medford2012scaling}, and the full set of noise spectra of Q1 at temperatures from $\SI{0.14}{\kelvin}$ to $\SI{1.2}{\kelvin}$.
    \textbf{f}, We examine the microwave effect on the qubit coherence time by applying the Hahn echo sequence on Q1. During the wait time, we apply a microwave signal far from the resonance of either qubit, to capture the incoherent noise induced. We measure $T_2^\mathrm{Hahn}$ varying the microwave power at $T=\SI{0.14}{\kelvin}$ and $T=\SI{1}{\kelvin}$. We observe a notably less evident effect from the microwave at $T=\SI{1}{\kelvin}$, compared to at $T=\SI{0.14}{\kelvin}$.
    }
    \label{fig:fESR_noise}
\end{figure*}

\begin{figure*}[ht]
    \centering
    \includegraphics[angle = 0]{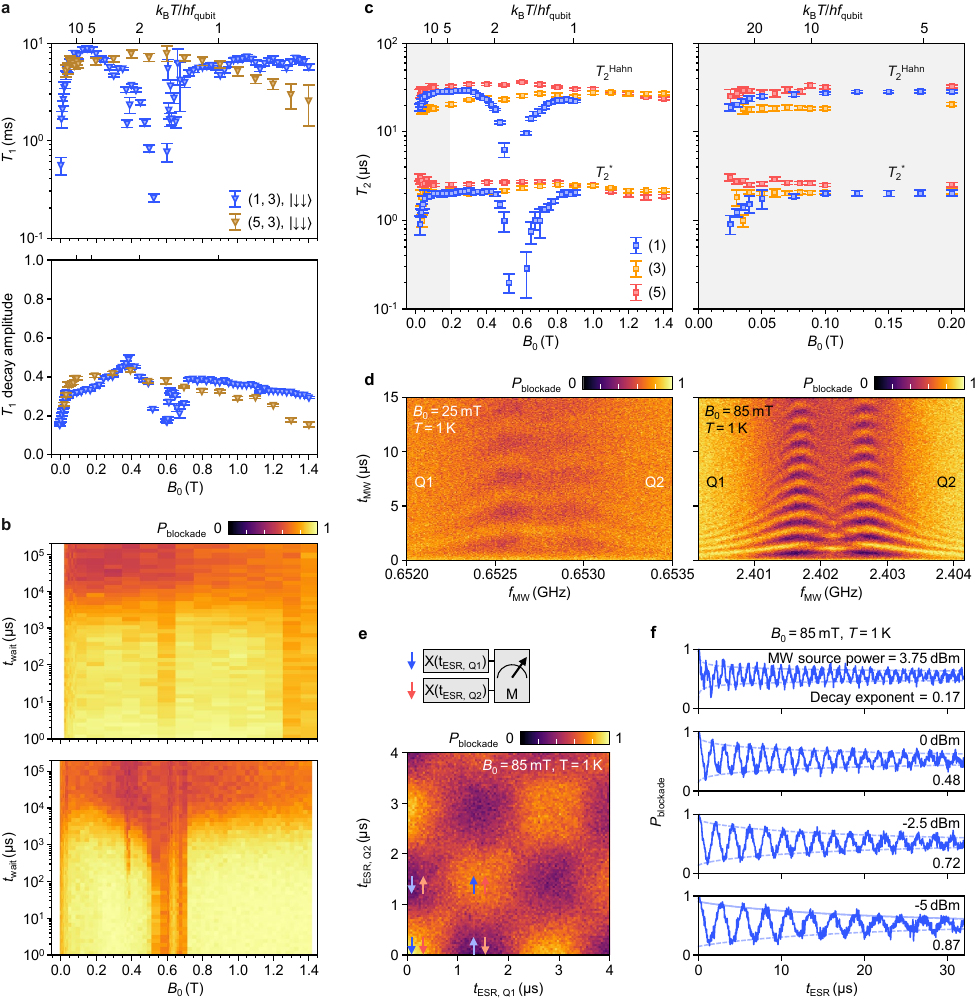}
    \caption{\textbf{$B_\mathrm{0}$ dependence.}
    \textbf{a}, $T_1$ and the decay amplitude as a function of $B_0$ at $T=\SI{1}{\kelvin}$. In $(1,3)$, $T_1$ exhibits a notable drop at low $B_0$ and near the hot spot induced by excited state crossings. The reduced decay amplitude is caused by the degraded spin readout around the hot spot, and additionally the small qubit energy relatively to the thermal energy at low $B_0$.
    \textbf{b}, Measured $T_1$ decay curves as a function of $B_0$ at $T=\SI{1}{\kelvin}$. The curves are fitted with the same method as described in Extended Data Fig.~\ref{fig:T1s_raw}~b.
    \textbf{c}, $T_2$ as a function of $B_0$ down to $\SI{25}{\milli\tesla}$ at $T=\SI{1}{\kelvin}$ in several charge configurations. $T_2^*$ and $T_2^\mathrm{Hahn}$ are almost $B_0$-invariant in the three- and five-electron configurations, but experience a drop around the hot spot in the one-electron configuration. The effect is highly local, and the qubit performance is consistent across configurations at low $B_0$ until $\SI{50}{\milli\tesla}$.
    \textbf{d}, Rabi oscillations in $(5,3)$ at ultra-low $B_0$ of $\SI{25}{\milli\tesla}$ and $\SI{85}{\milli\tesla}$, where the qubit energy is only $\SI{3.3}{\percent}$ and $\SI{11.4}{\percent}$ of the thermal energy. Due to the small $\Delta E_\mathrm{Z}$, crosstalk and deviation from the standard parity basis $\{\ket{\downarrow\downarrow}, \ket{\downarrow\uparrow}, \ket{\uparrow\downarrow}, \ket{\uparrow\uparrow}\}$ become significant.
    \textbf{e}, Simultaneously driven Rabi oscillations on both qubits, showing the alternation of the four parity basis states.
    \textbf{f}, Resonant Rabi oscillation of Q1 as a function of microwave power at $B_0 = \SI{85}{\milli\tesla}$ and $T=\SI{1}{\kelvin}$. The decay envelops are fitted to $ae^{-(t/T_2^\mathrm{Rabi})^c}+d$, where $a$ and $c$ are the decay amplitude and exponent, and $d$ is around 0.5. In general, we notice a reduction in the decay exponent at lower $B_0$, especially with faster driving. Possible causes are off-resonance driving on the ancilla qubit, decoherence during off-resonance driving, or an enhanced effect from the microwave. The coherence does not appear to be affected, and the quality factor of the Rabi oscillation is improved with faster driving. 
    }
    \label{fig:B0_dependence}
\end{figure*}

\begin{figure*}
    \centering
    \includegraphics[angle = 0]{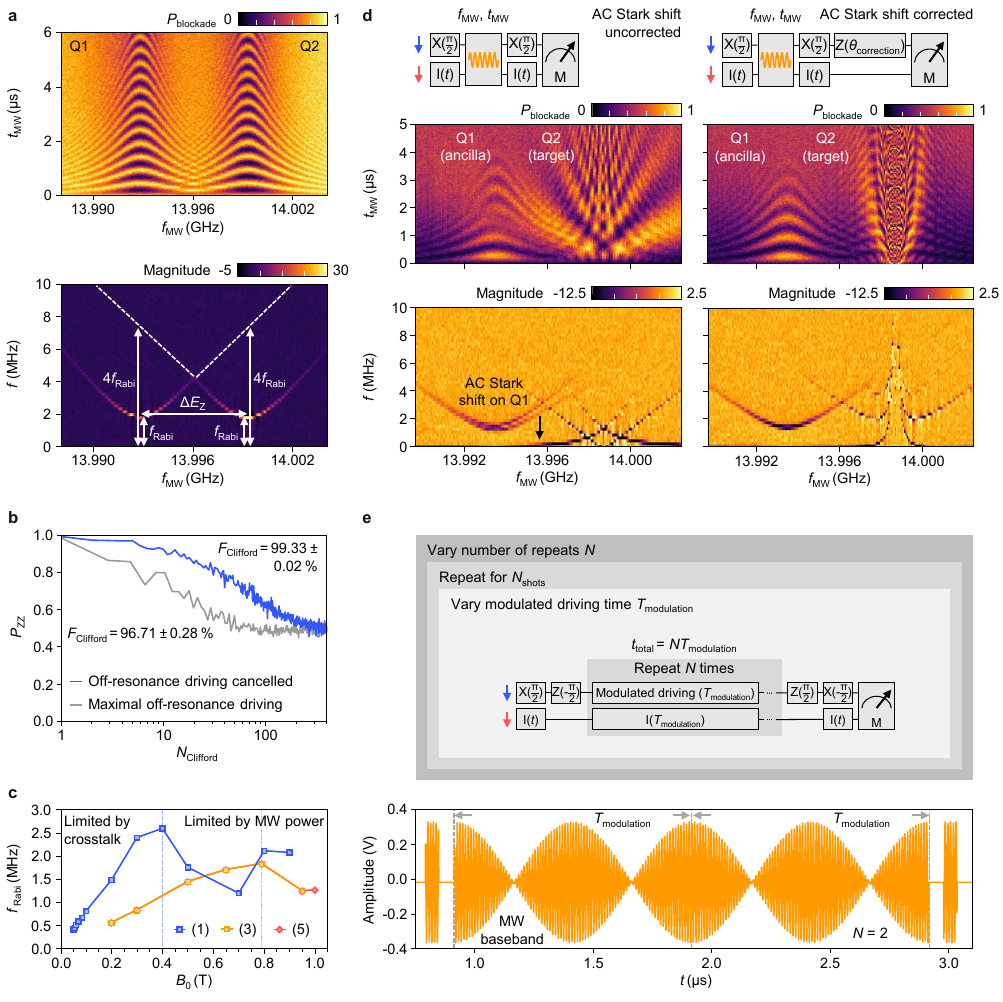}
    \caption{\textbf{Qubit crosstalk and the SMART protocol.}
    \textbf{a}, Crosstalk due to off-resonance driving at $B_0 = \SI{0.5}{\tesla}$ and $T=\SI{1}{\kelvin}$, plotted in time and frequency domain. In this measurement, $\Delta E_\mathrm{Z}$ and microwave power are set such that when Q1 is resonantly driven at $f_\mathrm{Rabi}$, the Rabi frequency of the off-resonance driving on Q2 is exactly $4f_\mathrm{Rabi}$ to meet the cancellation condition in equation~(\ref{eq:off_res_driving}), with $N=4$. This also applies to the case where Q2 is resonantly driven and Q1 is off-resonantly driven.
    \textbf{b}, Single-qubit randomised benchmarking (RB) of Q1 with and without off-resonance driving at $B_0 = \SI{0.5}{\tesla}$ and $T=\SI{1}{\kelvin}$. We maximise and cancel the off-resonance driving using the relationship in a.
    \textbf{c}, $f_\mathrm{Rabi}$ used in single-qubit RB at different $B_0$. This is set to meet the off-resonance driving cancellation condition based on the $\Delta E_\mathrm{Z}$ in each $B_0$ and charge configuration following equation~(\ref{eq:off_res_driving}). We use $N=4$ for fast driving until we reach the limit of the microwave source at high $B_0$, where the power transmission in the microwave line becomes much weaker.
    \textbf{d}, Sequence for probing the AC Stark shift and the results in time and frequency domain, taken at $B_0 = \SI{0.5}{\tesla}$ and $T=\SI{1}{\kelvin}$. We use Q1 as the ancilla to probe the AC Stark shift. We prepare it in the $\mathrm{-Y}$ direction and apply a microwave pulse with varying frequency $f_\mathrm{MW}$ and duration $t_\mathrm{MW}$. We show the results with and without correction. Without correction, AC Stark shift is seen as the linear fringes that crosses the driven Rabi chevron of Q1. This will translate into coherent Z errors during two-qubit operation. 
    \textbf{e}, Sequence for the SMART protocol~\cite{hansen2022implementation}. The sequence prepares the target qubit along the $\mathrm{+X}$ axis, and drive it with a cosine-modulated microwave pulse for a duration of $T_\mathrm{modulation}$. The qubit is then projected back onto the $\mathrm{+Z}$ axis for measurement.
    }
    \label{fig:crosstalk}
\end{figure*}

\begin{figure*}
    \centering
    \hspace{-0.6cm}
    \includegraphics[angle = 0]{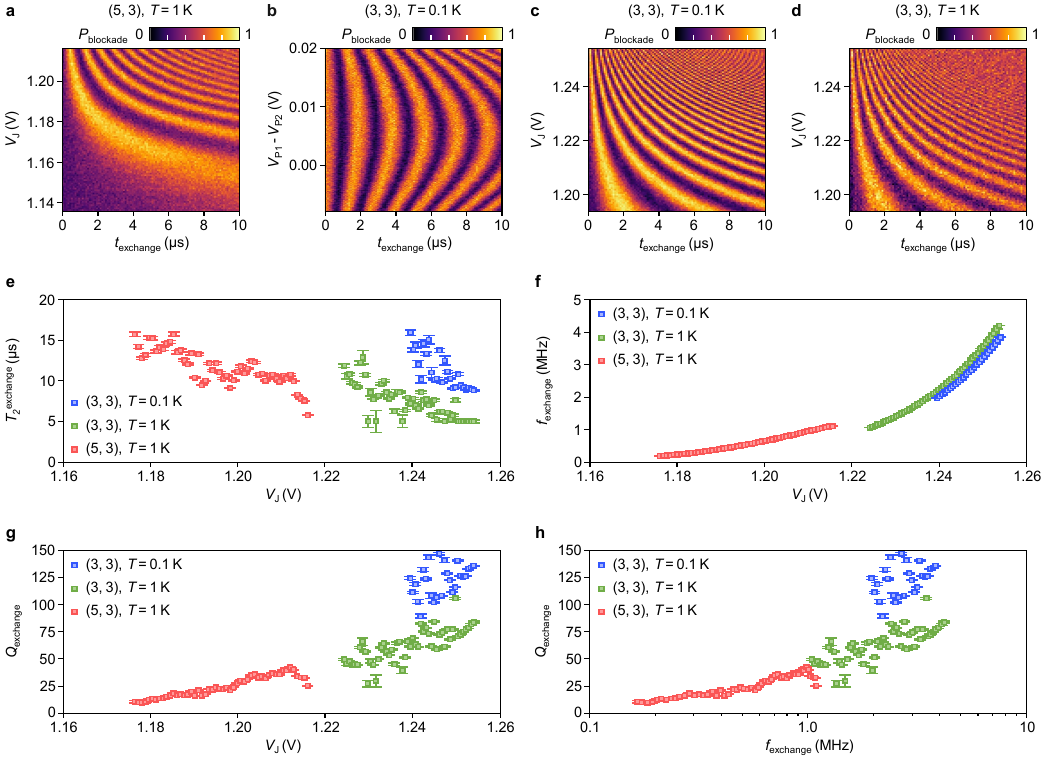}
    \caption{\textbf{Tuning of DCZ oscillations.}
    \textbf{a}, DCZ oscillations in $(5,3)$ as a function of time and $V_\mathrm{J}$ at $B_0 = \SI{0.79}{\tesla}$ and $T=\SI{1}{\kelvin}$.
    \textbf{b}, DCZ oscillations in $(3,3)$ as a function of time and $V_\mathrm{P1}-V_\mathrm{P2}$ at $B_0 = \SI{0.79}{\tesla}$ and $T=\SI{1}{\kelvin}$, showing the symmetric operation point.
    \textbf{c, d}, DCZ oscillations in $(3,3)$ as a function of time and $V_\mathrm{J}$ at $B_0 = \SI{0.79}{\tesla}$ and $T=\SI{0.1}{\kelvin}$ and $\SI{1}{\kelvin}$.
    \textbf{e}, $T_2$ of the DCZ oscillations $T_2^\mathrm{exchange}$ as a function of $V_\mathrm{J}$ at $B_0 = \SI{0.79}{\tesla}$, $T=\SI{1}{\kelvin}$. Error bars represent $\pm\SI{5}{\percent}$ fitting errors on individual oscillations.
    \textbf{f}, Frequency of the DCZ oscillations $f_\mathrm{exchange}$ as a function of $V_\mathrm{J}$ at $B_0 = \SI{0.79}{\tesla}$, $T=\SI{1}{\kelvin}$. Error bars represent $\pm\SI{5}{\percent}$ fitting errors on individual oscillations.
    \textbf{g}, Quality factor of the DCZ oscillations $Q_\mathrm{exchange}$ as a function of $V_\mathrm{J}$ at $B_0 = \SI{0.79}{\tesla}$, $T=\SI{1}{\kelvin}$, indicating that $f_\mathrm{exchange}$ at higher $V_\mathrm{J}$ outpaces $T_2^\mathrm{exchange}$. Error bars represent $\pm\SI{5}{\percent}$ fitting errors on individual oscillations.
    \textbf{h}, Quality factor of the DCZ oscillations $Q_\mathrm{exchange}$ as a function of $f_\mathrm{exchange}$ at $B_0 = \SI{0.79}{\tesla}$, $T=\SI{1}{\kelvin}$. Error bars represent $\pm\SI{5}{\percent}$ fitting errors on individual oscillations.
    }
    \label{fig:DCZ_raw}
\end{figure*}

\begin{figure*}
    \centering
    \includegraphics[angle = 0]{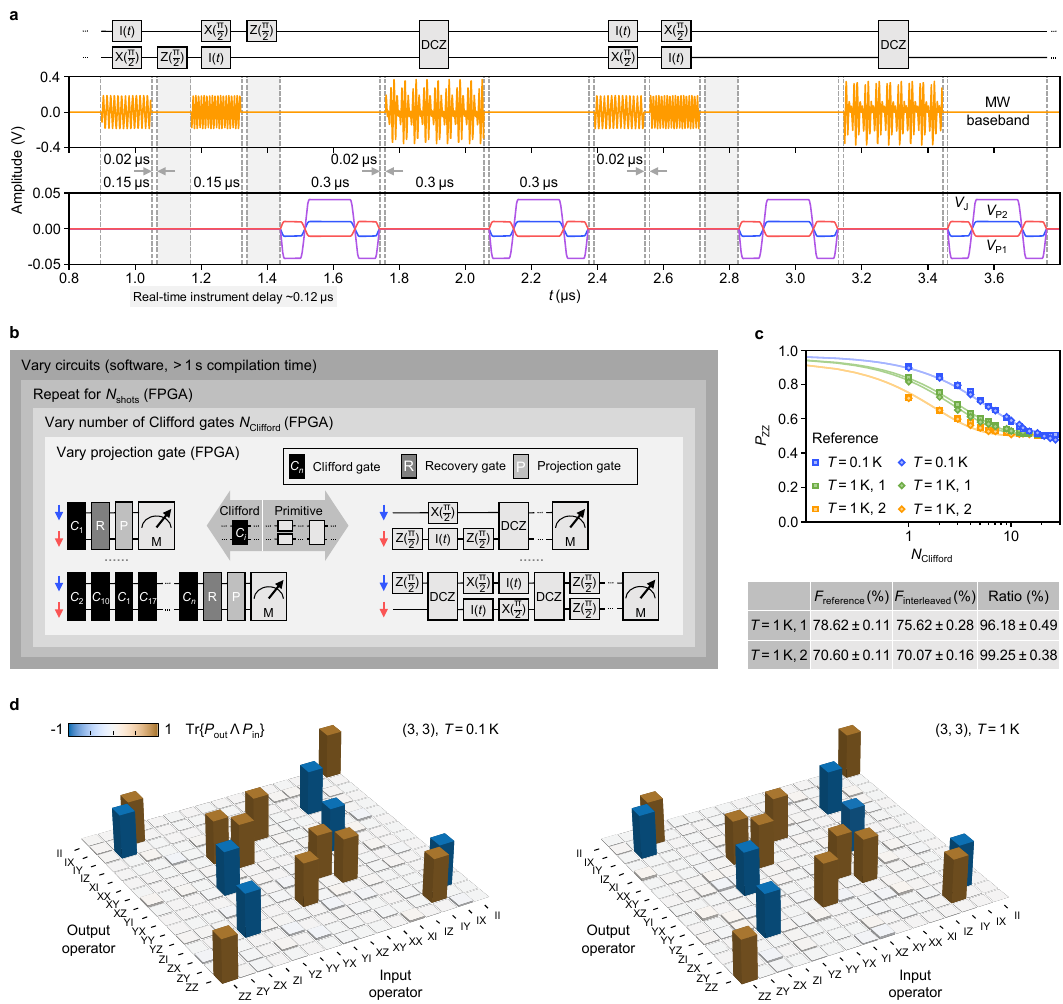}
    \caption{\textbf{Benchmarking and tomography of universal two-qubit logics.}
    \textbf{a}, An example random sequence with the microwave and voltage pulses generated by the FPGA. A DCZ gate includes two voltage pulses separated by an echoing two-tone $\mathrm{X}(\pi)$ pulse. The voltage pulse shape is designed to cancel any slow drift, and compensation is applied to P1, P2 while exchange is being pulsed. We set a padding of $\SI{0.02}{\micro\second}$ between two adjacent pulses. The real-time logics required for the FPGA to apply the sequences incur unintended gaps in the order of $\SI{0.1}{\micro\second}$ between some of the quantum gates. This introduces both coherent and incoherent errors. Ensuing efforts should target the minimisation of real-time logics and accurate synthesis of waveforms prior to the sequence run.
    \textbf{b}, The experiment and analysis protocols for two-qubit randomised benchmarking and FBT. The experimental gate sequences consist of random Clifford gates $C_i$ in the two-qubit space with a recovery gate $\mathrm{R}$ at the end. We then perform a projection $\mathrm{P}$ in $\mathrm{+ZZ}$ (no operation before parity readout) projection and $\mathrm{-ZZ}$ ($\pi$ pulse on a single qubit before parity readout).
    \textbf{c}, IRB results at $B_0=\SI{0.79}{\tesla}$, $T=\SI{0.1}{\kelvin}$ and $\SI{1}{\kelvin}$.
    \textbf{d}, Pauli transfer matrices (PTMs) for the DCZ gate at $B_0=\SI{0.79}{\tesla}$, $T=\SI{0.1}{\kelvin}$ and $\SI{1}{\kelvin}$, determined by FBT.
    }
    \label{fig:RB_and_FBT_raw}
\end{figure*}

\clearpage
\section*{Supplementary information}

\setcounter{figure}{0}
\captionsetup[figure]{name={\bf{Supplementary Fig.}},labelsep=line,justification=centerlast,font=small}

\begin{figure*}[!ht]
    \centering
    \includegraphics[angle = 0]{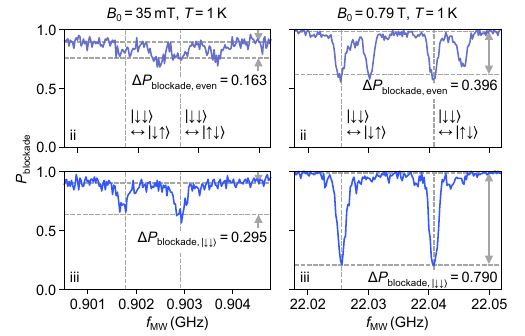}
    \caption{\textbf{Extraction of the algorithmic $\ket{\downarrow\downarrow}$ initialisation fidelity from the ESR spectra with exchange on.}
    \textbf{a}, When exchange is on, the ESR transitions provide information about the two-qubit state composition, as labelled in Fig.~\ref{fig:main_fig_2}~a. The first transition and the third transition from the left arise from the state transitions $\ket{\downarrow\downarrow}\leftrightarrow\ket{\downarrow\uparrow}$ and $\ket{\downarrow\downarrow}\leftrightarrow\ket{\uparrow\downarrow}$, and signify a $\ket{\downarrow\downarrow}$ state. With ideal spin inversion and readout, the amplitude of these two transitions yields an estimate of the fidelity of $\ket{\downarrow\downarrow}$ initialisation. The ESR spectra are measured by applying a microwave pulse at various frequencies and $V_\mathrm{J}$. In order to fully invert the spins, the pulse duration is calibrated to be $t_{\mathrm{X}_1(\pi)}$ at the single-qubit operation point. As the driving mechanism becomes different when exchange is on, this calibrated pulse does not fully invert the spins in these regimes. We first extract $\Delta P_\mathrm{blockade, even}$, the transition amplitude measured after Stage II of the algorithmic initialisation, which produces a mixed even-parity state, and $\Delta P_{\mathrm{initialisation},\ket{\downarrow\downarrow}}$, the transition amplitude measured after Stage III of the algorithmic initialisation, which produces a $\ket{\downarrow\downarrow}$ state. The initialisation fidelity is given by $F_{\mathrm{initialisation},\ket{\downarrow\downarrow}}=\Delta P_{\mathrm{initialisation},\ket{\downarrow\downarrow}}/(\Delta P_\mathrm{blockade, even}/0.5)$. We obtain $F_{\mathrm{initialisation},\ket{\downarrow\downarrow}}=\SI{90.77}{\percent}$ and $F_{\mathrm{initialisation},\ket{\downarrow\downarrow}}=\SI{99.56}{\percent}$ from the results at $B_0=\SI{35}{\milli\tesla}$ and $B_0=\SI{0.79}{\tesla}$. See Supplementary Fig.~\ref{fig:ESR_vs_J} for the full ESR spectra.
    }
    \label{fig:downdown_amplitude_extraction}
\end{figure*}

\begin{figure*}[!ht]
    \centering
    \includegraphics[angle = 0]{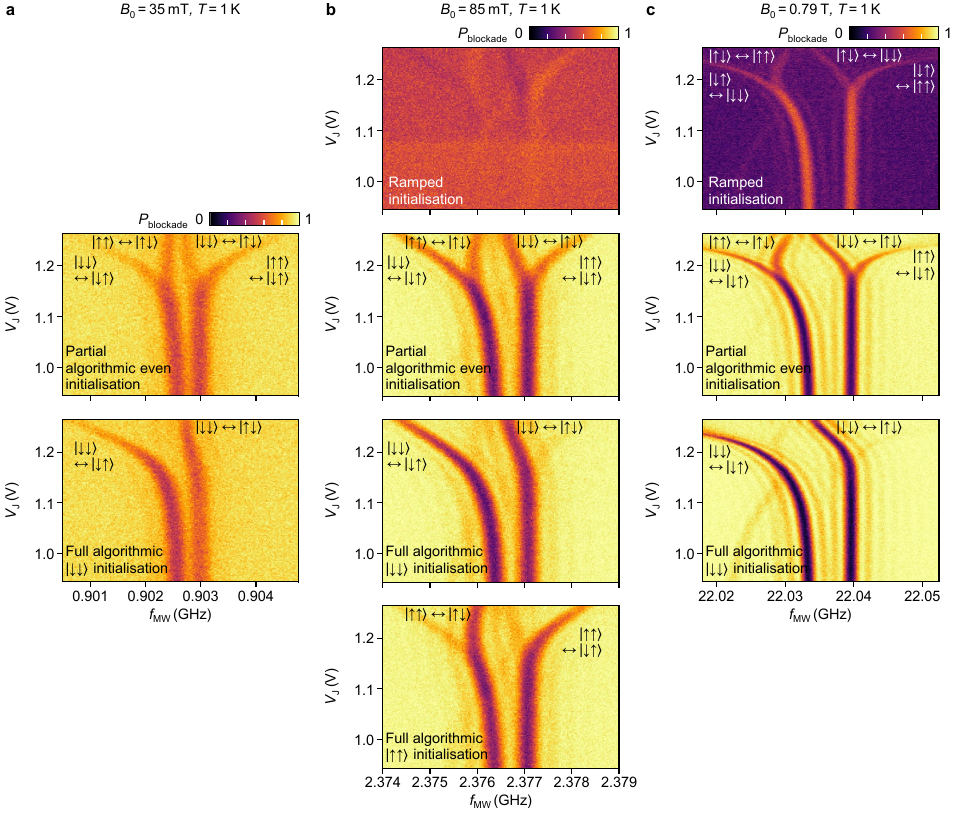}
    \caption{\textbf{ESR spectra as a function of $V_\mathrm{J}$ with different initialisation methods.}
    \textbf{a}, $B_0=\SI{35}{\milli\tesla}$ and $T=\SI{1}{\kelvin}$, where the thermal energy is 20 times greater than the qubit energies.
    \textbf{b}, $B_0=\SI{85}{\milli\tesla}$ and $T=\SI{1}{\kelvin}$, where the thermal energy is 8 times greater than the qubit energies.
    \textbf{c}, $B_0=\SI{0.79}{\tesla}$ and $T=\SI{1}{\kelvin}$, where the thermal energy is near the qubit energies.
    }
    \label{fig:ESR_vs_J}
\end{figure*}

\begin{figure*}[!ht]
    \centering
    \includegraphics[angle = 0]{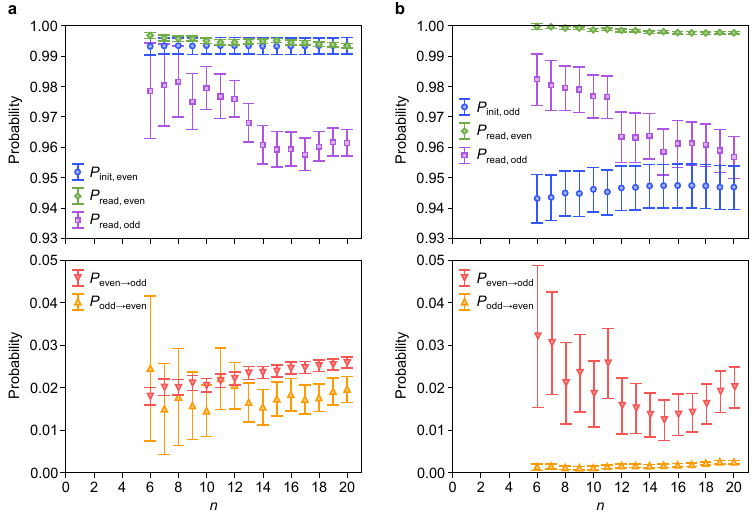}
    \caption{\textbf{Full results from SPAM analysis with machine learning.}
    \textbf{a}, All SPAM probabilities inferred from repeated readout outcomes through machine learning, for algorithmic $\ket{\downarrow\downarrow}$ initialisation at $B_0=\SI{0.79}{\tesla}$ and $T=\SI{1}{\kelvin}$.
    \textbf{b}, All SPAM probabilities inferred from repeated readout outcomes through machine learning, for algorithmic $\ket{\uparrow\downarrow}$ initialisation at $B_0=\SI{0.79}{\tesla}$ and $T=\SI{1}{\kelvin}$.
    }
    \label{fig:SPAM_analysis}
\end{figure*}

\begin{figure*}[!ht]
    \centering
    \includegraphics[angle = 0]{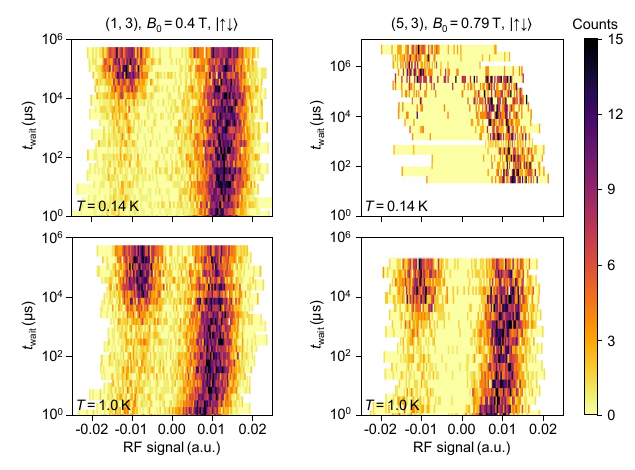}
    \caption{\textbf{FBT Estimated noise residual PTMs.}
    }
    Example readout histogram of long $T_1$ measurements.
    \label{fig:T1_histograms}
\end{figure*}

\begin{figure*}[!ht]
    \centering
    \includegraphics[angle = 0]{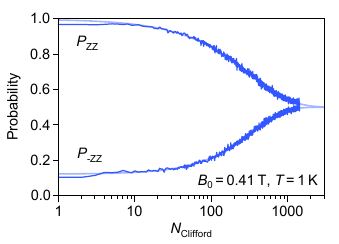}
    \caption{\textbf{$\mathrm{+Z}$ and $\mathrm{-Z}$ projection for a long single-qubit RB run.}
    We verify that with different projections, RB decays at about the same rate towards the same equilibrium of 0.5. The fitted Clifford fidelities from $P_\mathrm{Z}$, $P_\mathrm{-Z}$ and $P_\mathrm{Z} - P_\mathrm{-Z}$ are $99.8484\pm\SI{0.0021}{\percent}$, $99.8547\pm\SI{0.0016}{\percent}$ and $99.8519\pm\SI{0.0013}{\percent}$
    }
    \label{fig:long_RB_dual_projection}
\end{figure*}

\begin{figure*}[!ht]
    \centering
    \includegraphics[angle = 0]{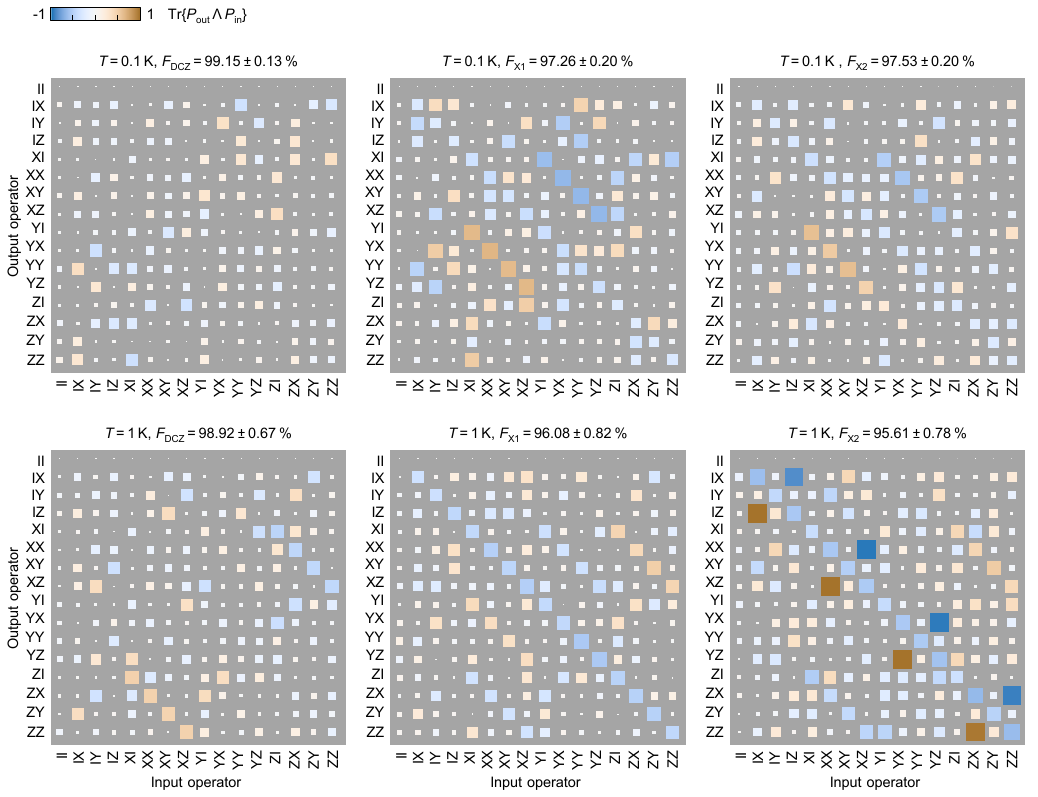}
    \caption{\textbf{FBT Estimated noise residual PTMs.}
    Projected error residual process matrices for the DCZ, $\mathrm{X}_1(\pi/2)$ and $\mathrm{X}_2(\pi/2)$ gates at $B_0=\SI{0.79}{\tesla}$, $T=\SI{0.1}{\kelvin}$ and $\SI{1}{\kelvin}$, extracted by FBT. We also show the consequent process fidelity for each gate.
    }
    \label{fig:error_generators}
\end{figure*}

\begin{figure*}[!ht]
    \centering
    \includegraphics[angle = 0]{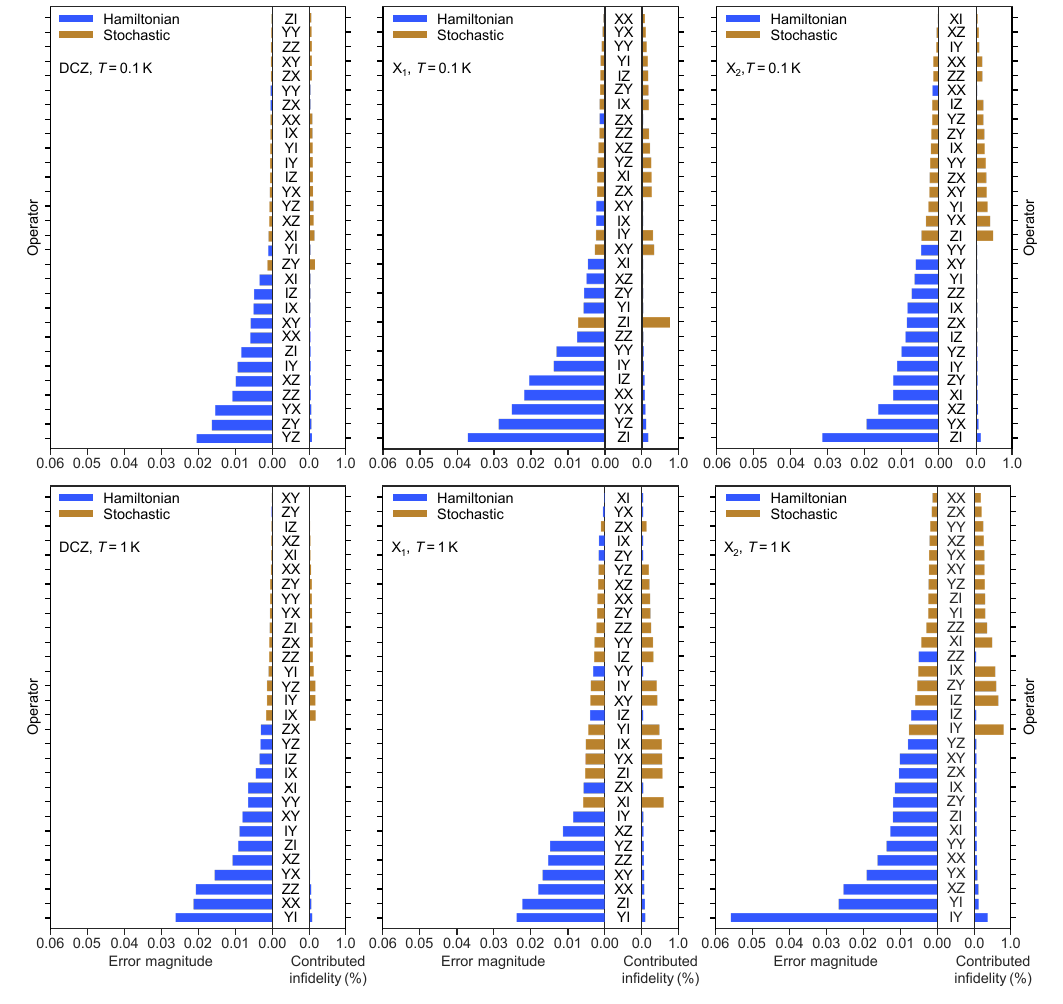}
    \caption{\textbf{All error channels obtained from the error generators.}
    The bar charts detail the magnitudes of both Hamiltonian and stochastic errors, as well as their contribution to entanglement infidelity for the $\mathrm{DCZ}$, $\mathrm{X}_1(\pi/2)$, and $\mathrm{X}_2(\pi/2)$ gates, at $B_0=\SI{0.79}{\tesla}$, $T=\SI{0.1}{\kelvin}$ and $\SI{1}\kelvin{}$.
    }
    \label{fig:error_channels}
\end{figure*}



\twocolumngrid
\clearpage
\bibliographystyle{naturemag}
\bibliography{main.bib}

\end{document}